\documentclass[11pt,letterpaper]{JHEP3}
\usepackage{graphics}
\usepackage{epsfig}
\usepackage{subfigure}
\usepackage{mathrsfs}
\usepackage{amsmath}
\usepackage{amsfonts}
\usepackage{amssymb}
\usepackage{graphicx}

\title{Holographic thermalization with a chemical potential in Gauss-Bonnet gravity}
\author{Xiao-Xiong Zeng\\
School of Science, Chongqing Jiaotong University, Chongqing, 400074, China\\
 \email{xxzeng@mail.bnu.edu.cn}}

\author{Xian-Ming Liu\\
Department of Physics, Hubei University for Nationalities, Enshi, 445000,
Hubei, China\\
\email{liuxianming1980@163.com}
}

\author{Wen-Biao Liu\\
Department of Physics, Beijing Normal University, Beijing, 100875, China\\
\email{wbliu@bnu.edu.cn}
}
 \abstract{Holographic thermalization is studied in the framework of
 Einstein-Maxwell-Gauss-Bonnet gravity.
 We use the  two-point correlation function and expectation value of Wilson loop,
  which are dual to the renormalized geodesic length and minimal  area surface in the bulk, to probe the thermalization.
The numeric result shows that larger the
 Gauss-Bonnet coefficient is,  shorter the  thermalization time is, and  larger the charge is,
  longer the thermalization time is, which implies that the Gauss-Bonnet coefficient
  can accelerate the thermalization  while the  charge  has an opposite effect.
 In addition, we obtain the functions with respect to the thermalization time for both the thermalization probes at a fixed charge
  and  Gauss-Bonnet coefficient, and on the basis of these functions, we obtain the thermalization
  velocity, which shows that the thermalization process is  non-monotonic. At the middle and later periods of the
   thermalization process, we find
  that there  is a phase transition point, which divides the
   thermalization into
   an acceleration phase and a deceleration phase. We also study the effect of the charge and Gauss-Bonnet coefficient
    on the phase transition point.

}

\preprint{}
\begin{document}
\bibliographystyle{ieeetr}
\bibliography{reference.bib}

\begin{thebibliography}{100}

\bibitem{Maldacena1998} J. M. Maldacena, The Large N limit of superconformal field theories and supergravity,
Adv. Theor. Math. Phys.2, 231 (1998) [hep-th/9711200].

\bibitem{Sonner2012} J. Sonner, A. G. Green, Hawking Radiation and Non-equilibrium Quantum Critical Current Noise,
 Phys. Rev. Lett.  109, 091601 (2012).

\bibitem{li030} W. J. Li, Y. Tian, H. Zhang, Periodically Driven Holographic Superconductor, JHEP 07, 030(2013) [arXiv:1305.1600 [hep-th]].

\bibitem{Murata2010} K. Murata, S. Kinoshita, N. Tanahashi, Non-equilibrium Condensation Process in a Holographic Superconductor, JHEP 1007, 050 (2010)  [arXiv:1005.0633[hep-th]].

\bibitem{Mukhopadhyay2013} A. Mukhopadhyay, Non-equilibrium fluctuation-dissipation relation from holography, Phys. Rev. D 87, 066004 (2013)   [arXiv:1206.3311[hep-th]].

\bibitem{Arnab}  K. Arnab, K. Sandipan, Steady-state Physics, Effective Temperature Dynamics in Holography, arXiv:1307.6607[hep-ph].

\bibitem{Shin2012}  N. Shin, Nonequilibrium Phase Transitions and a Nonequilibrium Critical Point from Anti-de Sitter Space and Conformal Field Theory Correspondence, Phys. Rev. Lett.   109, 120602 (2012)  [arXiv:1204.1971[hep-th]].


\bibitem{Baier2001} R. Baier, A. H. Mueller, D. Schiff, D. Son, Bottom up thermalization in heavy ion
collisions, Phys. Lett. B  502, 51 (2001) [hep-ph/0009237].


\bibitem{Garfinkle84} D. Garfinkle and L. A. Pando Zayas, Rapid Thermalization in Field Theory from
Gravitational Collapse, Phys. Rev. D 84,  066006 (2011) [arXiv:1106.2339 [hep-th]].

\bibitem{Garfnkle1202} D. Garfinkle, L. A. Pando Zayas and D. Reichmann, on Field Theory Thermalization
from Gravitational Collapse, JHEP 1202, 119  (2012) [arXiv:1110.5823 [hep-th]].

\bibitem{Allais1201} A. Allais and E. Tonni, holographic evolution of the mutual information, JHEP 1201
 102 (2012) [arXiv:1110.1607 [hep-th]].

\bibitem{Das343}S. R. Das, Holographic Quantum Quench, J. Phys. Conf. Ser. 343, 012027 (2012)
[arXiv:1111.7275 [hep-th]].

\bibitem{Steineder}D. Steineder, S. A. Stricker and A. Vuorinen, probing the pattern of holographic
thermalization with photons, arXiv:1304.3404 [hep-ph].

\bibitem{Wu1210} B. Wu, on holographic thermalization and gravitational collapse of massless scalar
fields,JHEP 1210, 133  (2012)[arXiv:1208.1393 [hep-th]].

\bibitem{Gao} X. Gao, A. M. Garcia-Garcia, H. B. Zeng, H. Q. Zhang, Lack of thermalization in holographic superconductivity,  arXiv:1212.1049 [hep-th].

\bibitem{Buchel2013} A. Buchel, L. Lehner, R. C. Myers and A. van Niekerk, Quantum quenches of holographic plasmas, JHEP 1305, 067 (2013) [arXiv:1302.2924[hep-th]].

\bibitem{Keranen2012}    V. Keranen, E. Keski-Vakkuri,  L. Thorlacius, Thermalization and entanglement
following a non-relativistic holographic quench, Phys. Rev. D 85,  026005 (2012)
[arXiv:1110.5035[hep-th]].


 \bibitem{Balasubramanian1} V. Balasubramanian {\it et al.}, Thermalization of Strongly Coupled Field
Theories, Phys. Rev. Lett. 106, 191601 (2011) [arXiv:1012.4753 [hep-th]].

\bibitem{Balasubramanian2} V. Balasubramanian {\it et al.}, Holographic Thermalization, Phys. Rev. D 84, 026010 (2011) [arXiv:1103.2683 [hep-th]].

\bibitem{GS}D. Galante and M. Schvellinger, Thermalization with a chemical potential from AdS
spaces, JHEP 1207, 096 (2012)  [arXiv:1205.1548 [hep-th]].

\bibitem{CK}E. Caceres and A. Kundu,    Holographic Thermalization with Chemical Potential, JHEP 1209, 055 (2012) [arXiv:1205.2354 [hep-th]].

\bibitem{Yang1} E. Caceres, A. Kundu, D. L. Yang, Jet Quenching and Holographic Thermalization with a Chemical Potential, arXiv:1212.5728 [hep-th].

\bibitem{Zeng2013}  X. X. Zeng and W. Liu, Holographic thermalization in Gauss-Bonnet gravity,  Phys. Lett. B 726,   481 (2013)
[arXiv:1305.4841[hep-th]].

\bibitem{Baron} W. H. Baron and M. Schvellinger, Quantum corrections to dynamical holographic
thermalization: entanglement entropy and other non-local observables,  arXiv:1305.2237 [hep-th].

\bibitem{Li3764} Y. Z. Li, S. F. Wu, G. H. Yang, Gauss-Bonnet correction to Holographic thermalization: two-point functions, circular Wilson loops and entanglement entropy, arXiv:1309.3764 [hep-th].

\bibitem{Baron1212} W. Baron, Damian Galante and M. Schvellinger, Dynamics of holographic thermalization,  arXiv:1212.5234 [hep-th].

\bibitem{Arefeva} I. Arefeva, A. Bagrov, A. S. Koshelev,  Holographic Thermalization from Kerr-AdS,   arXiv:1305.3267 [hep-th].

\bibitem{Hubeny} V. E. Hubeny, M. Rangamani, E.Tonni,  Thermalization of Causal Holographic Information, arXiv:1302.0853 [hep-th].

\bibitem{Arefeva6041} I. Y. Arefeva, I. V. Volovich, On Holographic Thermalization and Dethermalization of Quark-Gluon Plasma, arXiv:1211.6041 [hep-th].

\bibitem{Balasubramanianeyal6066} V. Balasubramanian {\it et al.},  Thermalization of the spectral function in strongly coupled two dimensional conformal field theories,  arXiv:1212.6066 [hep-th].

\bibitem{Balasubramanian4} V. Balasubramanian {\it et al.}, Inhomogeneous holographic thermalization,  arXiv:1307.7086[hep-th].

\bibitem{Balasubramanian3} V. Balasubramanian {\it et al.},
Inhomogeneous Thermalization in Strongly Coupled Field Theories, arXiv:1307.1487[hep-th].

\bibitem{Candelas46}  P. Candelas, G. T. Horowitz, A. Strominger and E. Witten, Vacuum configurations for
superstrings, Nucl. Phys. B 258 (1985) 46.

\bibitem{Nojiri049}  S. Nojiri and S. D. Odintsov,  Brane-world cosmology in higher derivative gravity or
warped compactification in the next-to-leading order of AdS/CFT correspondence,
JHEP 07 (2000) 049.

\bibitem{Boer5781} J. de Boer, M. Kulaxizi, A. Parnachev, Holographic Entanglement Entropy in Lovelock Gravities,  JHEP 1107, 109 (2011)[arXiv:1101.5781].


\bibitem{Hung5813}  L. Y. Hung, R.C. Myers, M. Smolkin, On Holographic Entanglement Entropy and Higher Curvature Gravity, JHEP 1104 (2011) 025 [arXiv:1101.5813].


\bibitem{Gregory010} A. Bhattacharyya, A. Kaviraj, A. Sinha, Entanglement entropy in higher derivative holography, JHEP 1308, 012 (2013).


\bibitem{Dong5713} X. Dong, Holographic Entanglement Entropy for General Higher Derivative Gravity, arXiv:1310.5713 [hep-th].


\bibitem{Li0210} Y. Z. Li, S. F. Wu, Y. Q. Wang, G. H. Yang, Linear growth of entanglement entropy in holographic thermalization captured by horizon interiors and mutual information, JHEP 09, 057 (2013) [arXiv:1306.0210 [hep-th]].

\bibitem{Guo2682} W. Z. Guo, S. He, J. Tao, Note on Entanglement Temperature for Low Thermal Excited States in Higher Derivative Gravity, arXiv:1305.2682 [hep-th].

\bibitem{Bhattacharyya5748} A. Bhattacharyya, M. Sharma, A. Sinha, On generalized gravitational entropy, squashed cones
 and holography, arXiv:1308.5748 [hep-th].


\bibitem{Myers83} R. C. Myers, S. Sachdev, A.Singh, Holographic Quantum Critical Transport without Self-Duality,
 Phys. Rev. D 83,  066017 (2011)[arXiv:1010.0433 [hep-th]].

\bibitem{Anninos} D. Anninos, G. Pastras, Thermodynamics of the Maxwell-Gauss-Bonnet anti-de Sitter Black Hole with Higher Derivative
 Gauge Corrections, JHEP 0907, 030 (2009) [arXiv:0807.3478 [hep-th]].

\bibitem{Cai} R. G. Cai, Gauss-Bonnet black holes in AdS spaces, Phys. Rev. D 65, 084014 (2002) [arXiv:hep-th/0109133].

\bibitem{Cremonini045} S. Cremonini, K. Hanaki, J. T. Liu and P. Szepietowski, Black holes in five-dimensional gauged supergravity with higher derivatives,  JHEP 0912, 045 (2009) [arXiv:0812.3572 [hep-th]].


\bibitem{Nojiri}  D. Anninos and G. Pastras, Thermodynamics of the Maxwell-Gauss-Bonnet anti-de Sitter Black Hole with Higher Derivative Gauge Corrections, JHEP 0907, 030 (2009) [arXiv:0807.3478 [hep-th]].


\bibitem{Astefanesei1334} D. Astefanesei, N. Banerjee, S. Dutta,(Un)attractor black holes in higher derivative AdS gravity, JHEP 0811, 070 (2008) [arXiv:0806.1334 [hep-th]].


\bibitem{Camanho3160}  X. O. Camanho, J. D. Edelstein, Causality constraints in AdS/CFT from conformal collider physics and Gauss-Bonnet gravity,  arXiv:0911.3160[hep-th].


\bibitem{0912.1944} X. O. Camanho, J. D. Edelstein, Causality in AdS/CFT and Lovelock theory,  arXiv:0912.1944 [hep-th].


\bibitem{Buchel} A. Buchel {\it et al.}, Holographic GB gravity in arbitrary dimensions, JHEP 1003, 111 (2010) [arXiv:0911.4257 [hep-th]].


\bibitem{Ge051} X. H. Ge, Y. Ling, Y. Tian and X. N. Wu, Holographic RG flows and transport coefficients in
 Einstein-Gauss-Bonnet-Maxwell theory,  JHEP  1112, 051 (2012) [arXiv:1112.0627[hep-th]].

\bibitem{Dominguez73} A. E. Dominguez, Emanuel Gallo, Radiating black hole solutions in Einstein-Gauss-Bonnet gravity, Phys. Rev. D 73, 064018 (2006) 	 [arXiv:gr-qc/0512150].
 \bibitem{Kobayashi} T. Kobayashi, A Vaidya-type radiating solution in Einstein-Gauss-Bonnet gravity and its application to braneworld, Gen. Rel. Grav. 37, 1869 (2005) [arXiv:gr-qc/0504027].

\bibitem{Maeda} H. Maeda, Effects of Gauss-Bonnet term on the final fate of gravitational collapse, Class. Quant. Grav.  23, 2155 (2006)
[arXiv:gr-qc/0504028].

\bibitem{ECaceres} E. Caceres, A. Kundu, J. F. Pedraza, W. Tangarife, Strong Subadditivity, Null Energy Condition
 and Charged Black Holes, arXiv:1304.3398 [hep-th].

\bibitem{Balasubramanian61}    V. Balasubramanian and S. F. Ross, Holographic particle detection, Phys. Rev. D 61,
044007 (2000) [arXiv:hep-th/9906226].

\bibitem{Maldacena80}   J. M. Maldacena, Wilson loops in large N field theories, Phys. Rev. Lett. 80, 4859 (1998)
[arXiv:hep-th/9803002].

\bibitem{Myers2834} R. C. Myers, M. F. Paulos and A. Sinha, Holographic Hydrodynamics with a Chemical
Potential,  JHEP  0906, 006 (2009) [arXiv:0903.2834 [hep-th]].


\bibitem{Ling2013} Y. Ling, C. Niu, J. P. Wu and Z. Y. Xian, Holographic Lattice in Einstein-Maxwell-Dilaton Gravity, arXiv:1309.4580[hep-th].


\end{thebibliography}
\section{Introduction}

AdS/CFT correspondence \cite{Maldacena1998} has been proved to be an effective tool to deal with the strong coupled systems.
 Especially in recent years, this correspondence has also been used extensively in
 non-equilibrium strong coupled systems \cite{Sonner2012, li030, Murata2010, Mukhopadhyay2013, Arnab, Shin2012}.
  One
 of the biggest motivation to use  this
gauge gravity duality to study the  non-equilibrium strongly coupled systems is that
 the thermalization time of quark
gluon plasma produced in RHIC and LHC experiments predicted by the perturbation theory is longer than the experiment result.
The reason for this difference is that the perturbation theory \cite{Baier2001}  treats the thermalization process as
 a near-equilibrium process,
where the static black hole in the bulk is dual to the boundary
system in equilibrium with finite temperature.
However, the sudden injection of energy in the thermalization process is a
  far-from-equilibrium
behavior of strongly coupled systems, the  holographic
 bulk thus should be  a highly dynamical spacetime, which can be described as black hole formation or black hole merger.
Based on this paradigm, there have been many models
 to study the   far-from-equilibrium thermalization behaviors
  \cite{Garfinkle84, Garfnkle1202, Allais1201, Das343, Steineder, Wu1210, Gao, Buchel2013, Keranen2012} recently.
 Especially in \cite{Balasubramanian1, Balasubramanian2},  Balasubramanian et al. find that one can use the two-point
correlation function, Wilson loop, and entanglement entropy, which can further be evaluated in the saddle point
 approximation in terms
of geodesic, minimal surface, and minimal volume individually, to detect the thermalization, where the  initial state in the
conformal field theory is dual to the AdS boundary in a higher dimensional space time, the sudden injection of energy is dual to the
collapse of a thin shell of dust, and the final equilibrium state is dual to a static black brane.
It is found that the holographic thermalization always proceeds in a
top-down pattern, namely the UV modes thermalize firstly, followed by the IR modes \footnote{On the weakly coupled side classical
 calculations have shown
that the thermalization process is of the bottom-up type, i.e. low energetic modes reach
thermal equilibrium first \cite{Baier2001}.}.
 They also find that there is a slight delay in the onset of thermalization and the entanglement entropy thermalizes slowest,
  which sets a timescale for equilibration.  Later, such an investigation is generalized to the bulk
geometry  with  electrostatic potential  \cite{GS,CK, Yang1} and high curvature corrections  \cite{Zeng2013,Baron, Li3764} to see how the chemical potential
and correction parameter affect the thermalization time in the boundary field theory, other extensions
on this topic please see \cite{Baron1212, Arefeva, Hubeny,
 Arefeva6041}. Made by Balasubramanian et al., there are further some elegant extensions on holographic thermalization very recently.
Firstly the time-dependent spectral functions in conformal field is found to be tractable\cite{Balasubramanianeyal6066}. With it, many quantities of
interest in the thermalization process, e.g., the time dependence
of the occupation number of field modes in the boundary gauge theory, can be calculated. Secondly, the  holographic thermalization
is extended to the inhomogeneous case
 by considering an $AdS_4$
weak field inhomogeneous collapse \cite{Balasubramanian4, Balasubramanian3}.
It  is found that  the AdS description of
the early time evolution is well-matched by free streaming, and the stress tensor approaches
that of second order hydrodynamics near the end of the early time interval.

The purpose of this  paper is to investigate holographic thermalization in the bulk with   curvature corrections and a gauge potential.
 By holography,  curvature correction corresponds to $\frac{1}{N}$ or $\frac{1}{\lambda}$ correction \cite{Candelas46, Nojiri049}
  to the boundary
  field theory \footnote{According to the AdS/CFT correspondence, it is well known that  IIB string theory on AdS5 $\times S^5$ background is dual to $D=4 ~N=4 ~SU(N_c) $ super Yang-Mills theory. In the limits of large $N_c$
and
large 't Hooft coupling, the SYM theory is dual to IIb supergravity which is low energy effective theory of superstring theory. So the higher curvature terms in the bulk theory
can arise as next to  leading order corrections in the $1/N $(large N) expansion of the boundary CFTs in the strong 't Hooft coupling limit.},
 and the gauge potential corresponds to
  a chemical potential in the boundary field theory. In Einstein gravity, there have been some works to study the
  effect of the chemical potential on the thermalization \cite{GS,CK}. And there are also some works to study the effect of correction parameter of the
  high order curvature on the   thermalization \cite{Zeng2013,Baron, Li3764}. Here we want to explore whether the
  chemical potential and the correction parameter have the same effect
  on the thermalization time. If not, our model will provide theoretically a  wider range of the thermalization
   time.  To observe the thermalization process
    in the dual conformal field theory,  we take the
two-point correlation  function and expectation value of Wilson loop as thermalization probes to study the
thermalization behavior \footnote{Usually, one also can use the
entanglement entropy to detect the  thermalization process. Recently there have been many works to study
holographic entanglement entropy in higher
derivative gravitational theories \cite{Boer5781, Hung5813, Gregory010, Dong5713, Li0210, Guo2682, Bhattacharyya5748}.
 Especially in \cite{Li3764},
the author
found that the entanglement entropy has similar behavior as the other observables during the thermalization process.
For simplicity, we will not study   entanglement entropy in this paper.}. According to  the AdS/CFT correspondence,
 this process equals to probing the evolution of a shell of charged
 dust that interpolates between a pure AdS and a charged Gauss-Bonnet AdS black brane by making use of the
   renormalized geodesic length and minimal  area surface.
 Concretely we first study the motion profile of the geodesic and minimal area, and then the  renormalized geodesic length and minimal  area surface
 in the charged  Gauss-Bonnet  Vaidya AdS black brane. When we study the effect of the gauge potential on the thermalization process, the
 Gauss-Bonnet coefficient is fixed and when we are interested in the effect of  Gauss-Bonnet coefficient, the gauge potential  is fixed.
 Our result shows that  larger the Gauss-Bonnet coefficient is,  easier the dual boundary system
 thermalizes, while  larger the charge is, harder the dual boundary system
 thermalizes. That is, the Gauss-Bonnet coefficient have an opposite effect on the thermalization time
 compared with the gauge potential.
 In addition, we also obtain the analytical functions of the  renormalized geodesic length with respect to the thermalization  time
 as well as  the renormalized  minimal  area surface  with respect to the thermalization time.  Based on the functions, we get  the thermalization
 velocity. Our result shows that the velocity is negative at the initial time and  positive at the middle and later periods,
  which indicates that the thermalization process is non-monotonic. We find that there is a phase transition point
  for the thermalization velocity at a fixed $Q$ and $\alpha$,
   which divides the thermalization process  into
   an acceleration phase and a deceleration phase. Effect of $Q$ and $\alpha$ on the phase transition points is
   also studied.

The remainder of this paper is organized as follows. In the next section, we shall provide a brief review of the charged
 Vaidya AdS black brane in Gauss-Bonnet gravity. Then
the holographic setup for non-local observables will be explicitly constructed in Section 3. Especially, we show that
 the relation between the  renormalized length and the two-point correlation function is still valid in the Gauss-Bonnet gravity.
 Resorting to numerical calculation, we
 perform a systematic analysis of how the Gauss-Bonnet coefficient and chemical potential affect the thermalization
  time  and thermalization velocity in Section 4. We end
up with some discussions in the last section.

\section{Charged Vaidya AdS black branes in Gauss-Bonnet gravity}
The $D(D \geq 5)$ dimensional Einstein-Maxwell theory with a negative
cosmological constant and a Gauss-Bonnet term  \footnote{The action in Eq.(\ref{action}) is not a consistent
truncation of an effective higher derivative gravitational action, it is one
special case where the higher order corrections of the Maxwell field have
vanishing coefficients.
One also can take into account the  corrections to the Maxwell field such as  in \cite{Myers83, Anninos}.}
representing  a quadratic
curvature correction is given by
\begin{equation}
\label{action} I=\frac{1}{16 \pi G_D}\int_\mathcal{M}d^Dx
\sqrt{-g} \left(R-2 \Lambda-4\pi G_D F_{\mu\nu}F^{\mu\nu}+\alpha L_{GB} \right),
\end{equation}
where  $G_D$  is the D-dimensional gravitational constant, $R$ is the Ricci scalar, $\Lambda$ is the negative cosmological constant, $\alpha$
is the Gauss-Bonnet coefficient, and $F_{\mu\nu}=\partial_{\mu} A_{\nu}-\partial_{\nu} A_{\mu}$,
\begin{equation}
\label{LGB} L_{GB} =R^2-4R_{\mu \nu}R^{\mu \nu}+R_{\mu \nu \sigma
\tau}R^{\mu \nu \sigma \tau}.
\end{equation}
From the action in Eq.(\ref{action}),
 many exact solutions have been found \cite{Cai, Cremonini045, Nojiri, Astefanesei1334,
 Camanho3160, 0912.1944, Buchel, Ge051}, here we are interested in the D-dimensional charged black brane solution
\begin{eqnarray}
  ds^2=-F(r)dt^2+F(r)^{-1}dr^2+\frac{r^2}{l^2}dx^2_n, \label{metric}
\end{eqnarray}
where
\begin{eqnarray}
 F(r)=\frac{r^2}{2\tilde{\alpha}}[1-\sqrt{1-\frac{4\tilde{\alpha}}{l^2}(1-\frac{Ml^2}{r^{D-1}}+\frac{Q^2l^2}{r^{2D-4}})}],
\end{eqnarray}
in which  $\ell =\sqrt{- \frac{(D-1) (D-2)}{2\Lambda}}$,  $\tilde{\alpha}=(D-3)(D-4)\alpha$, M and Q are related to the ADM black hole mass
$M_0$ and charge
$Q_0$
as follows
\begin{eqnarray}
  &&M_0=\frac{(D-2)MV_n}{16\pi G_D},\nonumber\\
  &&Q_0^2=\frac{2\pi(D-2)(D-3)Q^2}{G_D},
\end{eqnarray}
 where $V_n$
is the volume of the unit radius sphere $S^{D-2}$.  The $U(1)$ gauge potential reads
\begin{eqnarray} \label{chemical}
  A_t=-\frac{Q}{4\pi(D-3)}(r_h^{3-D}-r^{3-D})=u(1-\frac{r_h^{D-3}}{r^{D-3}}),
\end{eqnarray}
in which $r_h$ is the event horizon radius that  is characterized by  $ F(r_h) = 0$ and $u$ is the electrostatic potential, which can be identified with the chemical
potential of the dual field theory. The Hawking temperature of the charged Gauss-Bonnet AdS black
brane reads
\begin{eqnarray}\label{Temperature}
  T=\frac{\partial_r F(r)}{4\pi}|_{r_h}=\frac{(D-1)r_h-(D-3)Q^2\ell^2r_h^{5-2D}}{4\pi \ell^2},
\end{eqnarray}
which can be viewed as the temperature of the dual conformal field theory on the AdS boundary.
On the other hand,
as $r$  approaches to infinity, one can see the above black brane metric changes into
\begin{equation}
ds^2\rightarrow \frac{r^2}{\ell^2_{eff}}(-dt^2+d\tilde{x}_n^2)+\frac{\ell^2_{eff}}{r^2}dr^2,
\end{equation}
where
\begin{equation}
\tilde{x}_n=\frac{\ell_{eff}}{\ell}x_n,\ell^2_{eff}=\frac{2\tilde{\alpha}}{1-\sqrt{1-\frac{4\tilde{\alpha}}{\ell^{2}}}}.
\end{equation}
Thus this black brane solution is asymptotically AdS with AdS radius $\ell_{eff}$.

To get a Vaidya type evolving black brane, we would like first to make the coordinate
transformation $z=\frac{\ell^2}{r}$, in this case the black brane metric in Eq.(\ref{metric}) can be cast into
\begin{equation}
ds^{2}=\frac{\ell^2}{z^2}[-F(z)dt^{2}+F^{-1}(z)dz^{2}+dx_n^2]\label{metric1},
\end{equation}
where
\begin{equation}
\label{f}
F(z)=\frac{\ell^2}{2\tilde{\alpha}}\left[1-\sqrt{1-\frac{4\tilde{\alpha}}{\ell^2}\left(1-M
z^{D-1}\ell^{4-2D}+Q^2z^{2D-4}\ell^{10-4D}\right)}\right].
\end{equation}
 Then by introducing the
Eddington-Finkelstein coordinate system, namely
\begin{equation}
dv=dt-\frac{1}{F(z)}dz,
\end{equation}
 Eq.(\ref{metric1})  changes into
\begin{equation}
ds^2=\frac{\ell^2}{z^2} \left[ - F(z) d{v}^2 - 2 dz\ dv +
dx_n^2 \right] \label{collapsing}.
\end{equation}
Now the charged Gauss-Bonnet Vaidya AdS  black brane can be obtained by
freeing the mass parameter and charge parameter as  arbitrary functions of $v$\cite{Dominguez73,Kobayashi,Maeda}.
This  black brane is a solution of the equations of motion of the total action
\begin{equation}
I_{total}=I+I_{matter},
\end{equation}
where $I_{matter}$ is the action for external matter fields.
 As one can show,
such a metric is sourced by the null dust with the energy momentum tensor flux and gauge flux \cite{Dominguez73,Kobayashi,Maeda}
\begin{eqnarray}
  &&8\pi G_D T_{\mu\nu}^{matter}=z^{D-2}(\frac{D-2}{2}\dot{M}(v)-(D-2)z^{D-3}Q(v)\dot{Q}(v))\delta_{\mu v}\delta_{\nu v},\nonumber\\
  &&8\pi G_D J^{\mu}_{matter}=\sqrt{\frac{(D-2)(D-3)}{2}}z^D \dot{Q}(v)\delta^{\mu z},
\end{eqnarray}
where the dot stands for derivative with respect to coordinate $v$,  $M(v)$ and  $Q(v)$ are
 the mass and charge of a collapsing black brane\footnote{For more information about the energy condition in the time
 dependent background, please see \cite{ECaceres} }. It is obvious that for the charged Vaidya AdS black branes in Gauss-Bonnet gravity,
 the energy-momentum tensor  depends on not only  $M(v)$  but also  $Q(v)$.
 In this case, Eq. (\ref{collapsing}) describes the collapse of a thin-shell of charged dust from the boundary toward
the bulk interior of asymptotically anti-de Sitter spaces as $M$ and  $Q$ are substituted  by $M(v)$ and  $Q(v)$.

\section{Holographic thermalization}
In this section, we are going to investigate the thermalization process of a class of strongly coupled system.
According to the AdS/CFT correspondence, the rapid injection of energy on the boundary corresponds to the collapse of a black
brane in the AdS space.
So to describe the
thermalization process holographically, one should choose the mass $M(v)$ and charge $Q(v)$
properly so that it can describe the evolution of the charged dust. It was found that this properties can be achieved by
setting
the mass parameter and charge parameter as $M(v)=M\theta({v})$, $Q(v)=Q\theta({v})$ \cite{GS,CK}, where  $\theta({v})$ is
the
step function. In this case,
in the limit $v\rightarrow
-\infty$, the background  corresponds to a pure AdS space while in
the limit $v\rightarrow \infty$,
 it corresponds to a charged Gauss-Bonnet AdS black brane.
For the convenience of  numerical calculations, $M(v)$  and $Q(v)$ are usually
chosen as the smooth functions
\begin{equation}
M(v) = \frac{M}{2} \left( 1 + \tanh \frac{v}{v_0} \right),
\end{equation}
\begin{equation}
Q(v) = \frac{Q}{2} \left( 1 + \tanh \frac{v}{v_0} \right),
\end{equation}
where $v_0$ represents a finite shell thickness.

Having the construction of a model that describes the thermalization
process on the dual conformal field theory, we have to choose a set of extended non-local
observables \footnote{Because the local observables such as the energy-momentum tensor and its
derivatives can not explore deviations from thermal equilibrium in detail though it  provides valuable
information about the applicability of viscous hydrodynamics. } in the bulk
which allow us to evaluate the evolution of the system.
In this paper,  we shall focus mainly on the two-point correlation
function at equal time and expectation value of rectangular space-like Wilson loop,  which in the bulk  correspond to
the  renormalized geodesic length and minimal  area surface respectively.
For simplicity but without loss of generality, we shall set  the unit such that $\ell=1$ and  $r_h = 1$ in the later discussions. In addition, from  Eq.(\ref{f}), we  know that the mass and the charge have the relation $M=1+Q^2$, in this case we only need to change the charge to adjust the mass parameter during the numerical process.

\subsection{Renormalized geodesic length}
The relation between the renormalized geodesic length and two-point correlation
function at equal time has been discussed extensively \cite{Balasubramanian2, Balasubramanian61}. Here we will give a
 review of the relation to check whether it  is still valid in the Gauss-Bonnet gravity.

Start with the action of a complex scalar field of mass $m$ in the bulk background gravitational field $g_{ab}$
\begin{equation}
S=-\frac{1}{2}\int d^{d+1}X\sqrt{-g}[g^{ab}\nabla_a\bar\phi\nabla_b\phi+m^2\bar{\phi}\phi],
\end{equation}
which gives rise to the bulk propagator
\begin{equation}
G(x,y)=\langle\bar{\phi}(x)\phi(y)\rangle=\langle x|\frac{1}{iH}|\rangle y=\int_0^\infty dT\langle x|e^{-iHT}|y\rangle,
\end{equation}
where $H=-\frac{1}{2}[\nabla_a\nabla^a-m^2]$ can be thought of as the Hamiltonian of a fictitious
 quantum mechanical model with $T$ the proper time \footnote{For the derivation of the  Hamiltonian, please see appendix.}. With this in mind, one can reformulate this quantum mechanical model in terms of path integral over the worldlines of a  massive particle. In particular, the above bulk propagator can be expressed as
\begin{eqnarray}
G(x,y)&=&\int_0^\infty dT\mathcal{D}Xe^{is}\nonumber\\
&=&\mathcal{N}\int_0^\infty dT\prod_\tau \int DX(\tau)\sqrt{-g(\tau)}e^{i\int_0^Td\tau[\frac{1}{2}(g_{\mu\nu}\frac{dX^\mu}{d\tau}\frac{dX^\nu}{d\tau}-m^2)+\frac{R}{6}]}\nonumber\\
&=&\mathcal{N}\int_0^\infty dT\prod_\tau \int DX(\tau)\sqrt{-g(\tau)}e^{i\int_0^1d\tau[\frac{1}{2}(T^{-1}g_{\mu\nu}\frac{dX^\mu}{d\tau}\frac{dX^\nu}{d\tau}-T\hat{m}^2)]}\nonumber\\
\end{eqnarray}
with $X(0)=y$,  $X(T)=x$, and $\hat{m}^2=m^2-\frac{R}{3}$. In the saddle point approximation, i.e.,
\begin{equation}
T=\frac{\sqrt{-g_{\mu\nu}\frac{dX^\mu}{d\tau}\frac{dX^\nu}{d\tau}}}{\hat{m}},
\end{equation}
the bulk propagator can be evaluated as
\begin{equation}
G(x,y)\propto e^{i\int_0^1d\tau(-\hat{m}\sqrt{-g_{\mu\nu}\frac{dX^\mu}{d\tau}\frac{dX^\nu}{d\tau}})}\label{geodestics},
\end{equation}
with $X^\mu(\tau)$ the classical trajectory satisfying the above equation
 of motion. Obviously  Eq.(\ref{geodestics}) is consistent with the formulation in \cite{Balasubramanian2, Balasubramanian61}
\begin{equation}
\langle {\cal{O}} (t_0,\textbf{x}) {\cal{O}}(t_0, \textbf{x}')\rangle  \approx
e^{-\Delta {\tilde{L}_{ren}}} ,\label{llll}
\end{equation}
 where $\Delta$  is the conformal dimension of scalar operator $\cal{O}$, which is similar to $\hat{m}$ in Eq.(\ref{geodestics}),
 and
$\tilde{L}_{ren}$ indicates the renormalized length of the bulk geodesic between the points  $(t_0,
x_n)$ and $(t_0,x_n')$  on the AdS boundary. In other words, the Gauss-Bonnet coefficient has not effect
 on the dual relation between the renormalized geodesic length
and the two-point correlation
function. In what follows, we will make use of  Eq.(\ref{llll}) to explore how the  Gauss-Bonnet coefficient and gauge potential affect the thermalization time.  The reason that  we use the   renormalized geodesic  length $\tilde{L}_{ren}$ in  Eq.(\ref{geodestics}) is  that
the geodesic  length is divergent near the boundary, which is related to the divergent part as
\begin{equation}\label{lren}
\tilde{L}_{ren}= \tilde{L} +2\ell_{eff}\ln z_0.
\end{equation}
where $ \tilde{L} =\sqrt{-dS^2}$ is the geodesic  length between the points $(t_0,
x_n)$ and $(t_0,x_n')$ on the AdS boundary and $z_0$ is the IR radial cut-off.
Next, we are concentrating on studying   $\tilde{L}$.
 Taking into account the spacetime symmetry of our Vaidya type black brane,
  we can simply let $(t_0,
x_n)$ and $(t_0,x_n')$  have identical
  coordinates except $x_1=-\ell_{eff}\frac{l}{2}\equiv-\frac{\tilde{l}}{2}$ and $x'_1=-\ell_{eff}\frac{l}{2}\equiv-\frac{\tilde{l}}{2}$
  with $\tilde{l}$ the separation between these two points on the boundary. In order to make
   the notation as simple as possible, we would like to rename this exceptional coordinate $x_1$ as $x$ and employ it to
    parameterize the trajectory such that the proper length in the charged Gauss-Bonnet Vaidya AdS  black brane can be given by

\begin{equation}
  \tilde{L}= \int_{-\frac{\tilde{l}}{2}}^\frac{\tilde{l}}{2} dx
\frac{\sqrt{1-2z'(x)v'(x) - F(v,z) v'(x)^2}}{z(x)} ,\label{false}
\end{equation}
where the prime denotes the derivative with respect to $x$ and
\begin{equation}
F(v,z)=\frac{1}{2\tilde{\alpha}}\left[1-\sqrt{1-4\tilde{\alpha}\left(1-M(v)
z^{D-1}+Q(v)^2z^{2D-4}\right)}\right].
\end{equation}
Note that the integrand in
Eq.(\ref{false}) can be thought of as the Lagrangian $\cal{L}$ of a fictitious system with $x$ the proper time.
Since the Lagrangian does not depend explicitly on $x$, there is an associated conserved quantity
\begin{equation} \label{h}
 {\cal{H}}  ={\cal{L}}-v'(x)\frac{\partial \cal{L}}{\partial v'(x) } -  z'(x)\frac{\partial \cal{L}}{\partial z'(x) }= \frac{1}{z(x) \sqrt{1-2z'(x)v'(x) -F (v,z) v'(x)^2} }.
\end{equation}
In addition, based on the
Lagrangian $\cal{L}$ of the system, we obtain
\begin{eqnarray}
\frac{\partial\cal{L}}{\partial z}&=&\frac{-v'(x)^2\partial _z F(z,v) \cal{H}}{2}-\frac{1}{z(x)^3 \cal{H} },\\
\frac{\partial} {\partial x}\frac{\partial\cal{L}}{\partial z'}&=&-v''(x)\cal{H}.
\end{eqnarray}
So
the equation of motion for $z(x)$  can be written as
\begin{eqnarray} \label{zgequation}
0 &=& 2 - 2 v'(x)^2 F(v,z) - 4 v'(x) z'(x) - 2 z(x) v''(x) + z(x)
v'(x)^2 \partial_z F(v,z).
\end{eqnarray}
Similarly,  the equation of motion for $v(x)$ can be solved as
\begin{eqnarray} \label{vgequation}
0&=&v'(x)z'(x)\partial_zF(v,z)+\frac{1}{2}v'(x)^2\partial_vF(v,z)+v''(x)F(v,z)+z''(x),
\end{eqnarray}
in which
\begin{eqnarray}
  &&\partial_z F(z,v)=\frac{(D-1)M(v)z^{D-2}-(2D-4)Q^2(v)z^{2D-5}}{\sqrt{1-4\tilde{\alpha}(1-M(v)z^{D-1}+Q^2(v)z^{2D-4})}},\nonumber\\
  &&\partial_v F(z,v)=\frac{M'(v)z^{D-1}-2Q(v)Q'(v)z^{2D-4}}{\sqrt{1-4\tilde{\alpha}(1-M(v)z^{D-1}+Q^2(v)z^{2D-4})}}.
\end{eqnarray}
Next, we turn to studying the equations of motion in Eq.(\ref{zgequation}) and Eq.(\ref{vgequation}). Considering the reflection symmetry of our geodesic, we will use the following initial conditions
\begin{equation}\label{initial}
z(0)=z_*,  v(0)=v_* , v'(0) =
z'(0) = 0.
\end{equation}
As $z(x)$ and  $v(x)$ are solved, we can get the IR radial cut-off and thermalization time by
the boundary conditions as follows
\begin{equation}\label{regularization}
z(\frac{\tilde{l}}{2})=z_0, v(\frac{\tilde{l}}{2})=t_0,
\end{equation}
In addition, with the help of Eq.(\ref{h}) and  Eq.(\ref{initial}),  the proper length of geodesic in Eq.(\ref{false}) can be simplified as
\begin{equation} \label{r2}
\tilde{L} = 2 \int_{0}^{\frac{\tilde{l}}{2}} dx \frac{z_*}{z(x)^2},
\end{equation}
which will be more convenient for us to  obtain the  renormalized geodesic  length.
\subsection{Minimal area surface}
In this section, we are going to study the minimal  area surface,  which in the dual conformal field theory   corresponds to
the  Wilson loop operator.
Wilson loop operator is defined as a path ordered integral of gauge field over a closed contour,
 and its expectation value is approximated geometrically  by the AdS/CFT correspondence as \cite{Balasubramanian2, Maldacena80}
\begin{equation} \label{area}
\langle W(C)\rangle \approx e^{-\frac{\tilde{A}_{ren}(\Sigma)}{2\pi\alpha'}},
\end{equation}
where $C$ is the closed contour, $\Sigma$ is the minimal bulk surface ending on $C$ with $\tilde{A}_{ren}$ its renormalized
minimal  area surface, and  $\alpha'$ is the Regge slope parameter. In the  Gauss-Bonnet gravity, we will assume  Eq.(\ref{area}) is still valid as that between the   renormalized geodesic length
and the two-point correlation
function.

Here we are focusing solely on the rectangular space-like Wilson loop. In this case, the enclosed rectangle can  always be
chosen to be centered at the coordinate origin and lying on the $x_1-x_2$ plane with the assumption that the corresponding
 bulk surface is invariant along the $x_2$ direction. This implies that the minimal  area surface   can be expressed as
\begin{equation}\label{lequation}
\tilde{A}=\int_{\frac{\tilde{l}}{2}}^{\frac{\tilde{l}}{2}}dx
\frac{\sqrt{1-2z'(x)v'(x) - F(v,z) v'(x)^2}}{z(x)^2},
\end{equation}
where we have set the separation along $x_2$ direction to be one and the separation along $x_1$ to be $\tilde{l}$ with $x_2$ renamed as $y$ and $x_1$ renamed as $x$.
As before,  from  Eq.(\ref{lequation}) we also can get a   Lagrangian $\cal{L}$ and with it we can find
 a conserved quantity, i.e.,
\begin{equation}
{\cal{H}}=\frac{1}{z(x)^2 \sqrt{1-2z'(x)v'(x) - F (v,z) v'(x)^2} },
\end{equation}
which can simplify our equations of motion as
\begin{eqnarray} \label{aequation}
0&=&4-4v'(x)^2F(v,z)-8v'(x)z'(x)-2z(x)v''(x)+z(x)v'(x)^2\partial_zF(v,z),\nonumber\\
0&=&v'(x)z'(x)\partial_zF(v,z)+\frac{1}{2}v'(x)^2\partial_vF(v,z)+v''(x)F(v,z)+z''(x).
\end{eqnarray}
Similarly, with the initial conditions as in (\ref{initial}) and the regularization cut-off as in (\ref{regularization}), the
renormalized minimal  area surface  can be cast into
\begin{equation}\label{aren}
\tilde{A}_{ren}=2\int_0^{\frac{\tilde{l}}{2}}dx
\frac{z^2_*}{z(x)^4}-\frac{2}{z_0}.
\end{equation}
Next, we will investigate the evolution of the renormalized minimal  area surface with respect to the thermalization time to
explore how the gauge potential and  Gauss-Bonnet coefficient affect the thermalization process.

\section{Numerical results}
In this section,  we will solve the equations of motion of
geodesic length and
minimal  area surface  numerically, and then explore how the chemical potential and  Gauss-Bonnet coefficient affect
 the thermalization time. Since there have been many works to study the effect of
the space time dimensions and boundary separation on the thermalization probes \cite{Balasubramanian1, Balasubramanian2, GS, CK},
 to avoid redundancy,
we mainly discuss the case $D=5$ and a fixed boundary separation in this paper. During the numerics, we will take the shell thickness and  UV
 cut-off as $v_0 = 0.01$, $z_0 = 0.01$ respectively.

According to the AdS/CFT correspondence, we know that the electromagnetic
field in the bulk is dual to the chemical potential in the dual quantum field theory, so we will use the  electromagnetic
field defined in Eq.(\ref{chemical}) to explore the effect of the chemical potential on the thermalization process in the AdS boundary.
 However, as stressed in \cite{GS, Myers2834, Ling2013},
 the chemical potential has energy
units in the dual field theory $([u]= 1/[L])$ while $A_{\mu}$
as defined in Eq.(\ref{action}) is dimensionless, thus one has to redefine the  electromagnetic
field as
$\tilde{A}_{\mu}=A_{\mu}/p$, where $p$
is a scale with length unit that depends on the particular compactification. In this case, $\tilde{A}_{\mu}$
 and  $u$  have the same unit and the chemical potential can be expressed as
\begin{equation}
u= \mathop{\lim}_{r\rightarrow\infty} \tilde{A}_{\mu}=\frac{Q r_h^{3-D}}{4 \pi p (D-3)}.
\end{equation}
In addition,  due to the conformal symmetry on the boundary, the quantity
which is physically meaningful is the ratio of
$u/T$ in asymptotically charged AdS space time, namely the chemical potential  measured with the
temperature as the unit. Therefore, through this paper we will use the following ratio
\begin{equation} \label{ratio}
\frac{u}{T}=\frac{Q}{p(D-3)[(D-1)r_h^{D-2}-(D-3)Q^2r_h^{3-D}]},
\end{equation}
to check the effect of the chemical potential on the thermalization time.
Obviously, for the case $r_h=1$,  Eq.(\ref{ratio}) shows that $u/T$ changes from $0\rightarrow \infty$
 provided $Q $ changes from $0\rightarrow \sqrt{\frac{D-1}{D-3}}$. In other words, to adjust the change of the ratio
 $u/T$ in all the range, we only need to change  $Q $  from $0\rightarrow \sqrt{\frac{D-1}{D-3}}$. For the case $D=5$ in
 this paper, we will choose $Q=0.00001, 0.5, 1$ in our numerical result.
 On the other hand, we will also consider the effect of the Gauss-Bonnet coefficient on the thermalization time, as in \cite{Zeng2013}
 we will take  $\alpha=-0.1, 0.0001, 0.08$  as the constraint
of
causality of dual field theory on the boundary is imposed\footnote{As  $\alpha$ is
dialed beyond the causality bound but within the Chern-Simons limit, it has not effect on the thermalization process. One can see the
 thermalization curves plotted in \cite{Li3764}.}.

\begin{figure}
\centering
\subfigure[$Q=0.00001, \alpha=-0.1$]{
\includegraphics[scale=0.55]{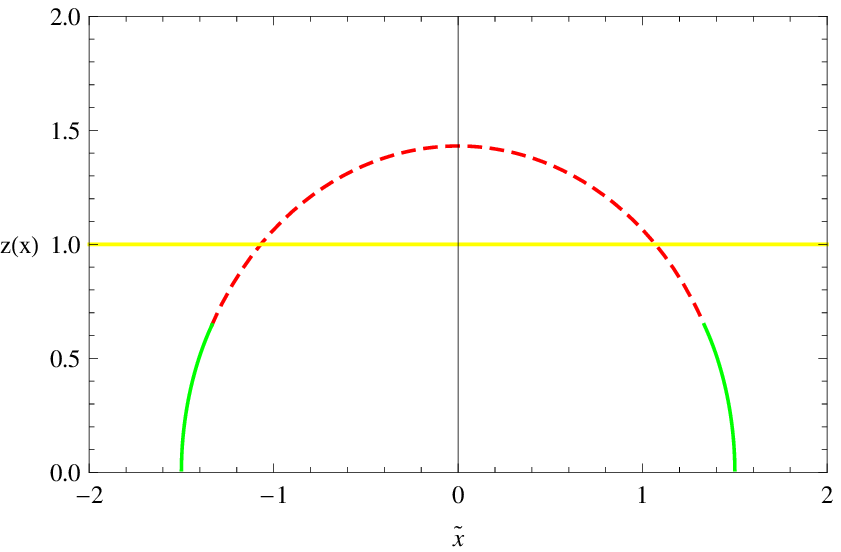}
\label{shell1_RN} }
\subfigure[$Q=0.00001, \alpha=0.0001$]{
\includegraphics[scale=0.55]{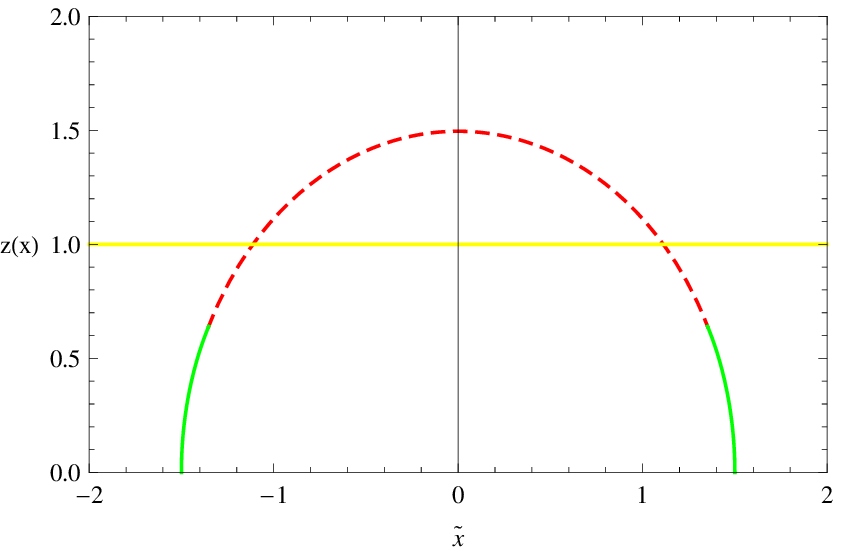}
\label{shell2_RN}
}
\subfigure[$Q=0.00001, \alpha=0.08$]{
\includegraphics[scale=0.55]{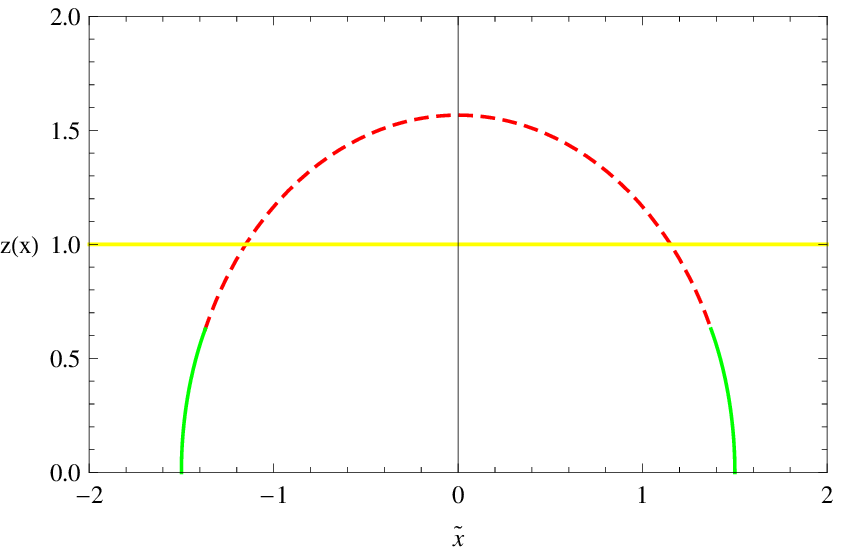}
\label{shell3_RN} }
\subfigure[$Q=0.5, \alpha=-0.1$]{
\includegraphics[scale=0.55]{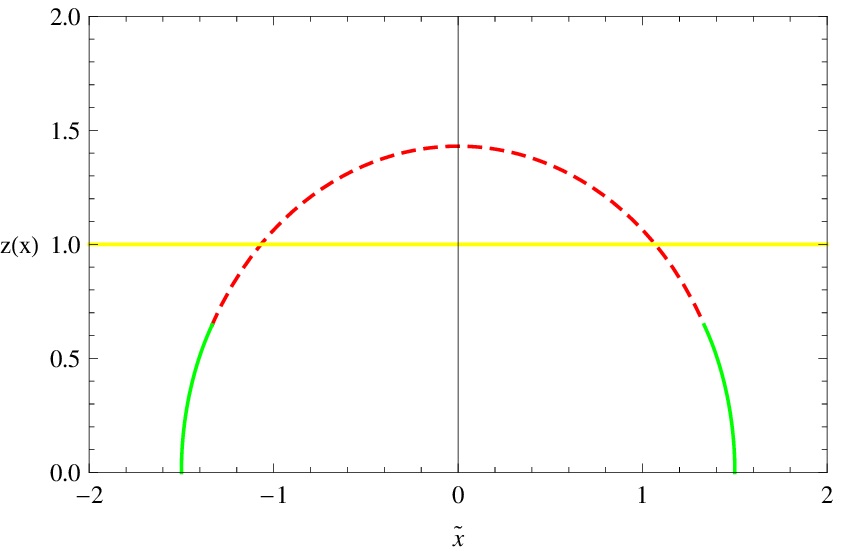}
\label{shell4_RN}
}
\subfigure[$Q=0.5, \alpha=0.0001$]{
\includegraphics[scale=0.55]{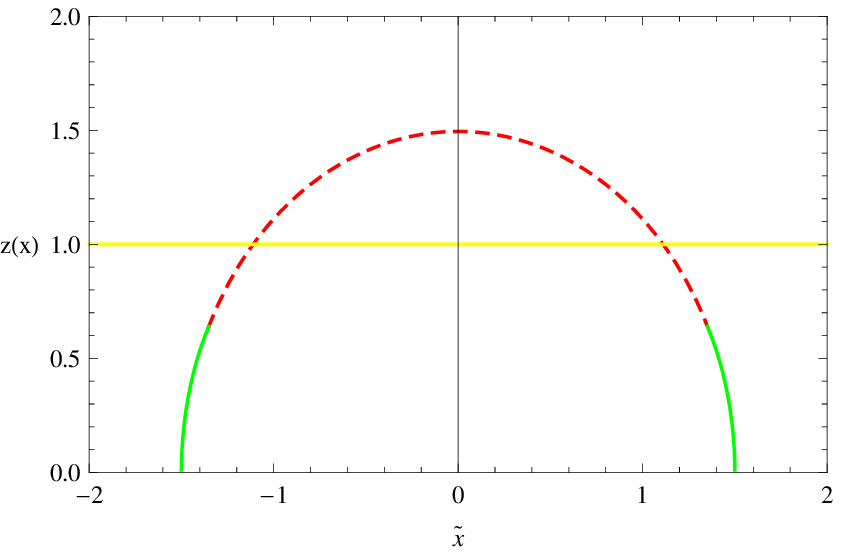}
\label{shell5_RN} }
\subfigure[$Q=0.5, \alpha=0.08$]{
\includegraphics[scale=0.55]{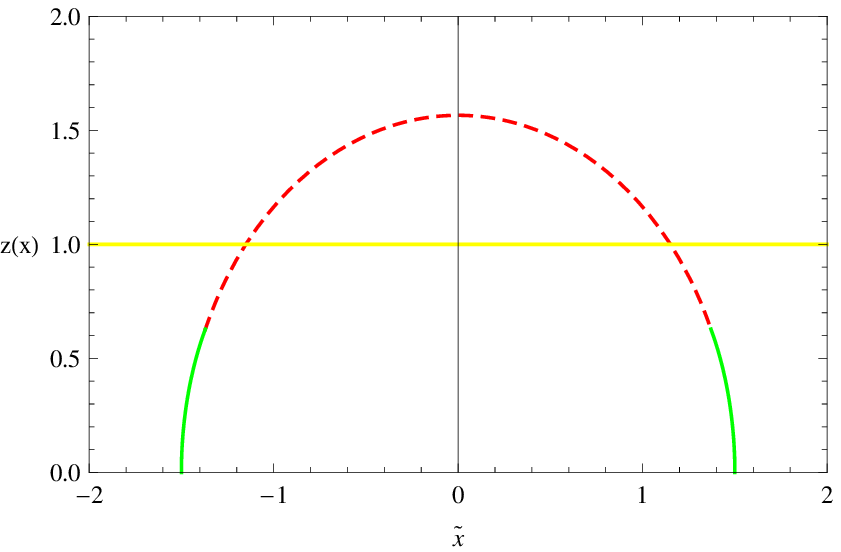}
\label{shell6_RN}
}
\subfigure[$Q=1, \alpha=-0.1$]{
\includegraphics[scale=0.55]{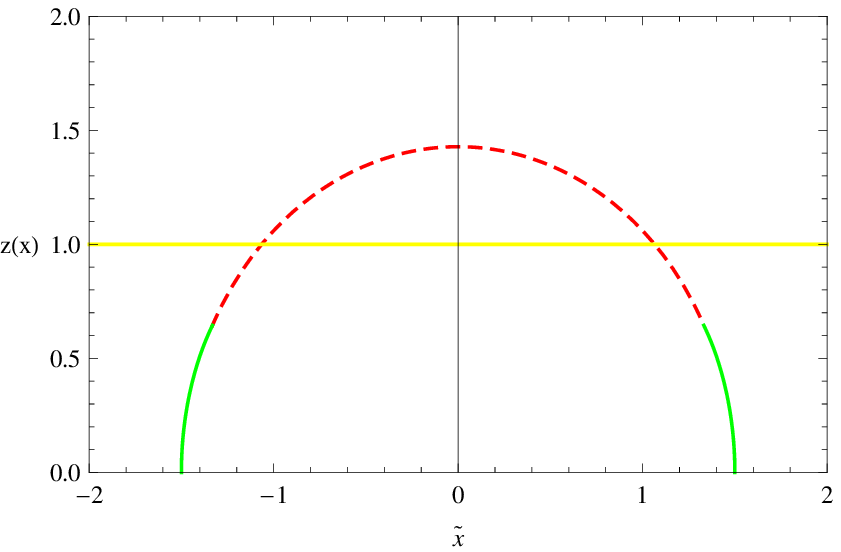}
\label{shell7RN} }
\subfigure[$Q=1, \alpha=0.0001$]{
\includegraphics[scale=0.55]{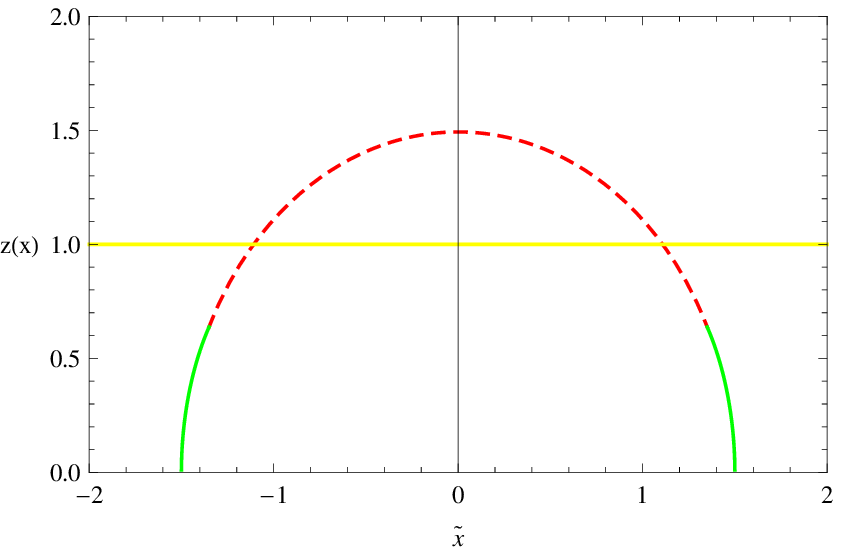}
\label{shell8RN} }
\subfigure[$Q=1, \alpha=0.08$]{
\includegraphics[scale=0.55]{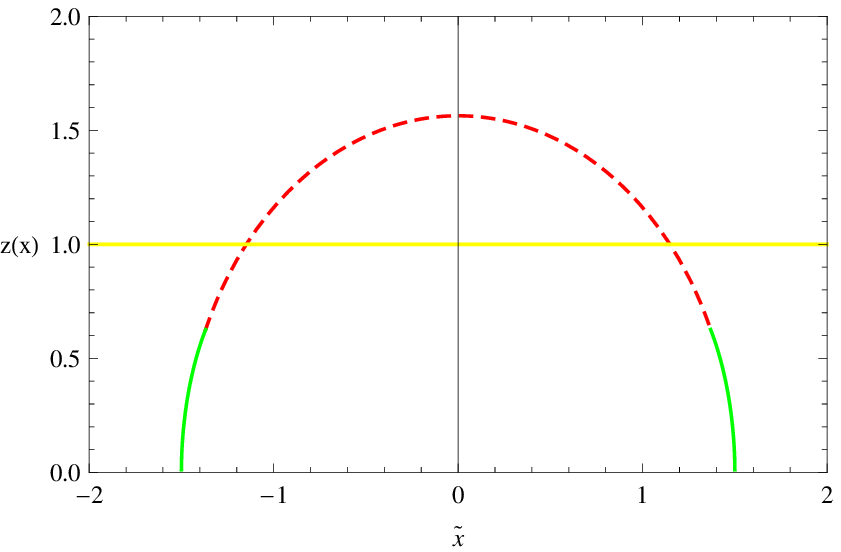}
\label{shell9RN} }
 \caption{\small Motion profile of the geodesics in the charged Gauss-Bonnet Vaidya AdS black brane. The separation of the
boundary field theory operator pair is $\tilde{\ell}=3$ and the initial time is $v_{\star}=-0.856$. The black brane
horizon is indicated by the yellow line. The  position of the shell is described by the junction between the dashed red line and the green line.} \label{fig1}
\end{figure}

\begin{figure}
\centering
\subfigure[$Q=0.00001, \alpha=-0.1$]{
\includegraphics[scale=0.55]{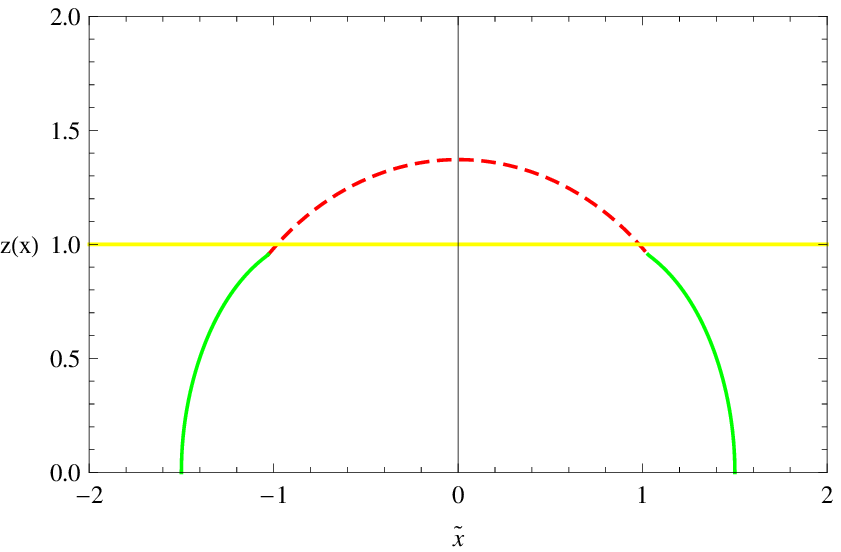}
\label{shell1_RN} }
\subfigure[$Q=0.00001, \alpha=0.0001$]{
\includegraphics[scale=0.55]{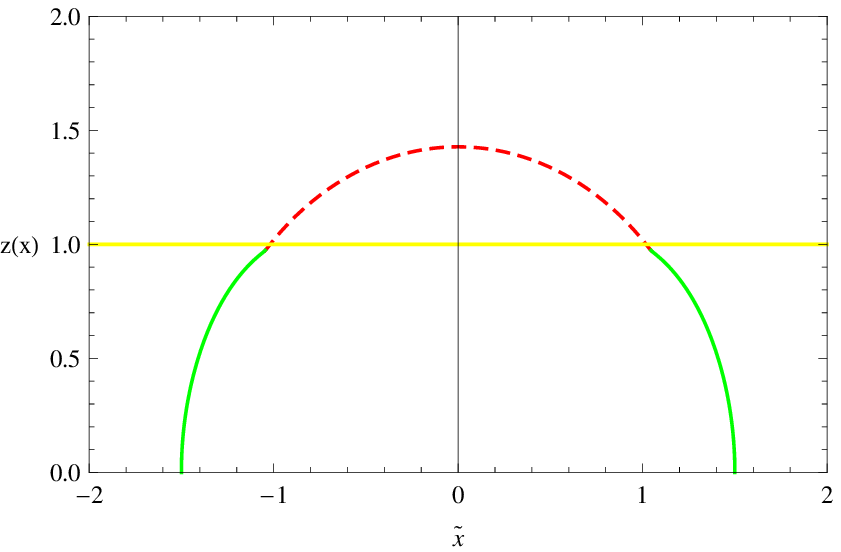}
\label{shell2_RN}
}
\subfigure[$Q=0.00001, \alpha=0.08$]{
\includegraphics[scale=0.55]{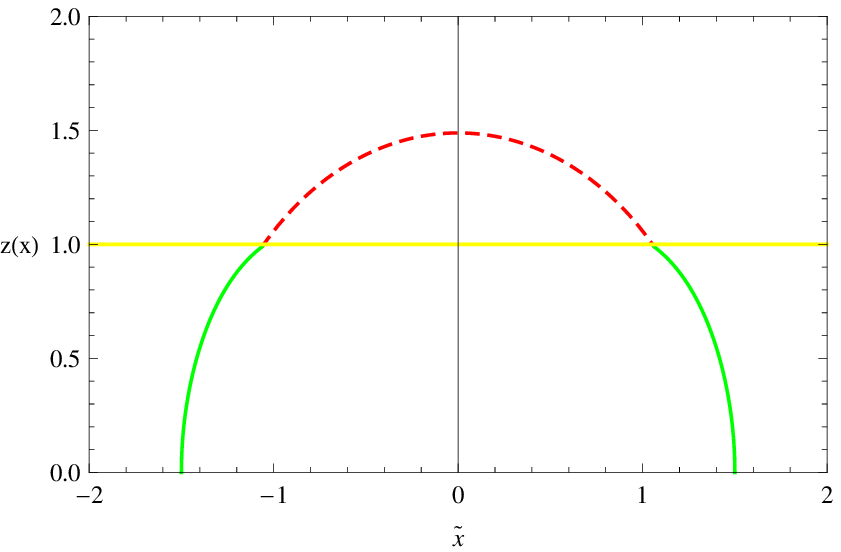}
\label{shell3_RN} }
\subfigure[$Q=0.5, \alpha=-0.1$]{
\includegraphics[scale=0.55]{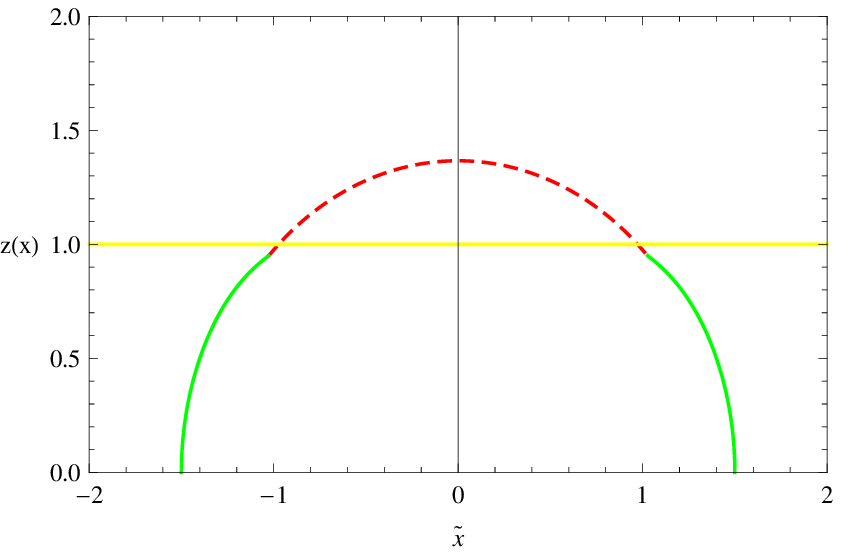}
\label{shell4_RN}
}
\subfigure[$Q=0.5, \alpha=0.0001$]{
\includegraphics[scale=0.55]{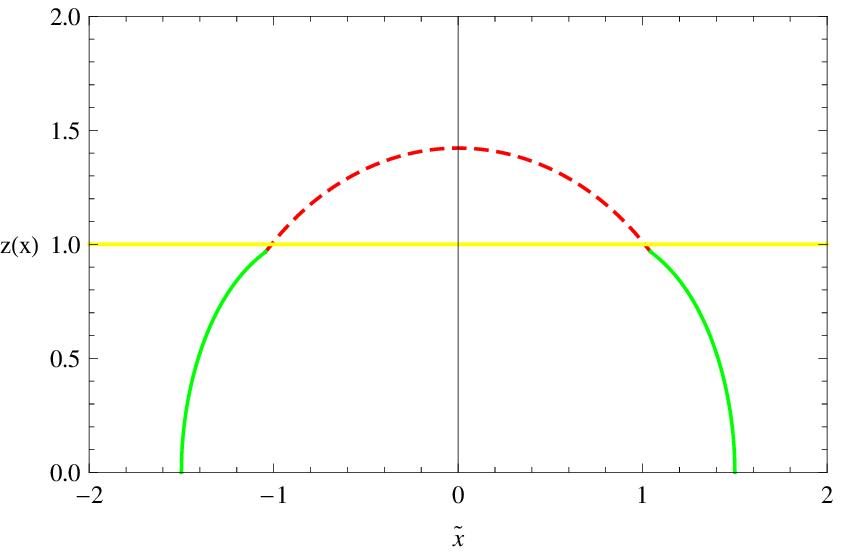}
\label{shell5_RN} }
\subfigure[$Q=0.5, \alpha=0.08$]{
\includegraphics[scale=0.55]{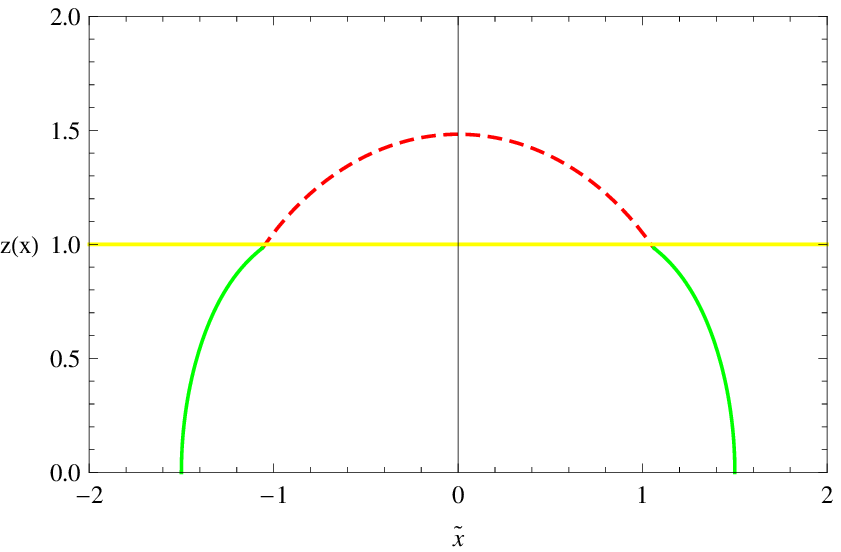}
\label{shell6_RN}
}
\subfigure[$Q=1, \alpha=-0.1$]{
\includegraphics[scale=0.55]{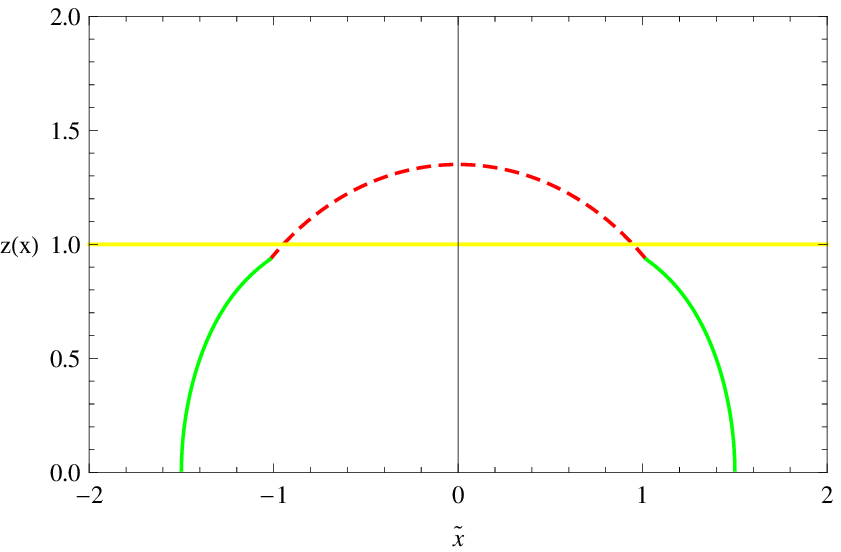}
\label{shell7RN} }
\subfigure[$Q=1, \alpha=0.0001$]{
\includegraphics[scale=0.55]{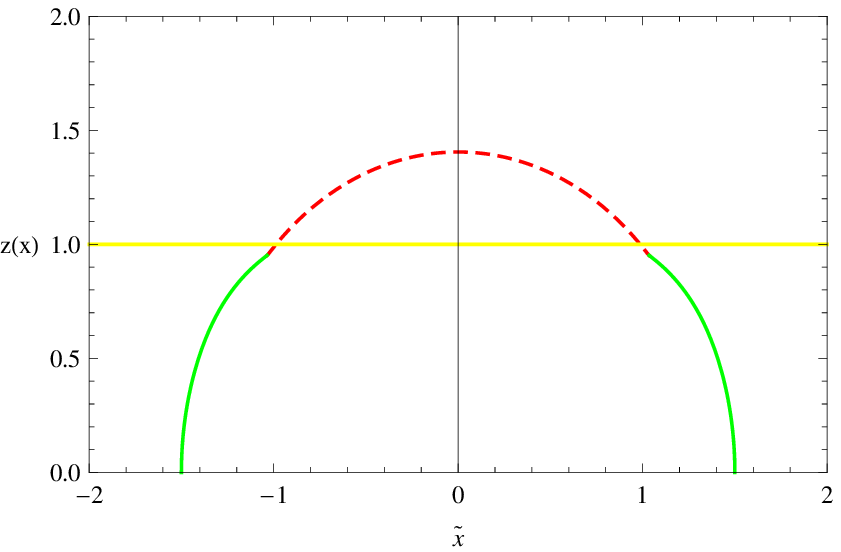}
\label{shell8RN} }
\subfigure[$Q=1, \alpha=0.08$]{
\includegraphics[scale=0.55]{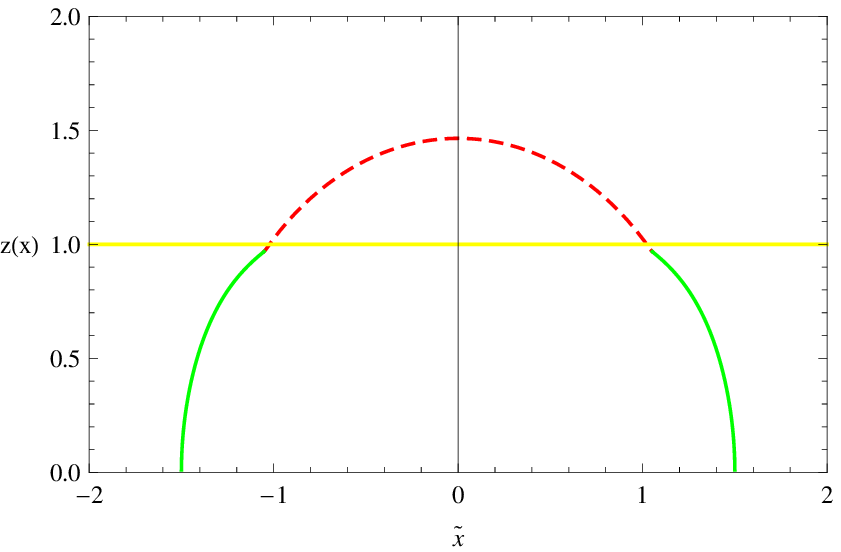}
\label{shell9RN} }
 \caption{\small Motion profile of the geodesics in the charged  Gauss-Bonnet Vaidya AdS black brane. The separation of the
boundary field theory operator pair is $\tilde{\ell}=3$ and the initial time is $v_{\star}=-0.456$.  The black brane
horizon is indicated by the yellow line. The  position of the shell is described by the junction between the dashed red line and the green line.} \label{fig11}
\end{figure}

 As the boundary conditions in Eq.(\ref{initial}) is adopted, the  equations of motion of the geodesic in Eq.(\ref{zgequation}) and  Eq.(\ref{vgequation}) for different  $\alpha$ and different $Q$ can be solved numerically. When we are interested in the
effect of  $Q$ on the motion profile of the  geodesic, the  Gauss-Bonnet coefficient  $\alpha$ is fixed, and when we are interested in the
effect of    Gauss-Bonnet coefficient  $\alpha$, the charge  $Q$ is fixed.
Since different initial time $v_{\star}$
corresponds to different stage of the motion of the geodesics, we  also  discuss the effect of   $v_{\star}$ on the motion profile. Figure (\ref{fig1})
and Figure (\ref{fig11}) plot  the  motion profile of the  geodesic at the initial time $v_{\star}=-0.856, -0.456$ respectively
for different charge and Gauss-Bonnet coefficient.  In both figures,
 the  horizontal direction is the motion profile of the geodesics for different Gauss-Bonnet coefficients while the
 vertical direction is the motion profile of the geodesics for different charges. From Figure (\ref{fig1})
and   Figure (\ref{fig11}), we know that as the initial time increases, the shell of the dust approaches to the horizon of the charged
  Gauss-Bonnet black brane, which means that for the larger initial time, the thermalization
   has been behaved longer. For different initial time, we also keep a watchful eye on how the  Gauss-Bonnet coefficient
    and charge affect the thermalization time, which are listed in Table (\ref{tab:g1}). From this table,
    we can observe that for both the two different
initial times, as the Gauss-Bonnet coefficient grows, the thermalization time  decreases,  which means that the  quark
gluon plasma in the dual conformal theory  is easier to be thermalized. But as the charge grows,
we find for different initial times, the thermalization time has different variation trends. For the case $v_{\star}=-0.856$,
 it is found that for a fixed Gauss-Bonnet coefficient, as the charge grows the thermalization time has a little difference,
 while for  $v_{\star}=-0.456$, the thermalization time increases as the charge grows. That is, the charge has little
 effect on the
 thermalization time at the initial stage of the thermalization.
 We also can observe the effect of the Gauss-Bonnet coefficient  and charge on the motion profile  of the geodesics.
From  Figure (\ref{fig11}), we find for a fixed charge, e.g. $Q=0.5$,  as the  Gauss-Bonnet coefficient
grows from   $\alpha=-0.1$ to  $\alpha=0.08$, the shell will approaches to the horizon of the black brane and
drops into the horizon lastly.  In other words, for  $\alpha=-0.1$, the  quark
gluon plasma  in the conformal field theory is thermalizing while for  $\alpha=0.08$, it is thermalized.
For a fixed Gauss-Bonnet coefficient, e.g. $\alpha=0.08$, we find at  $Q=0.00001$, the shell
 lies above the horizon while at  $Q=1$, the shell lies below the horizon.

\begin{figure}
\centering
\subfigure[$\alpha=-0.1$]{
\includegraphics[scale=0.55]{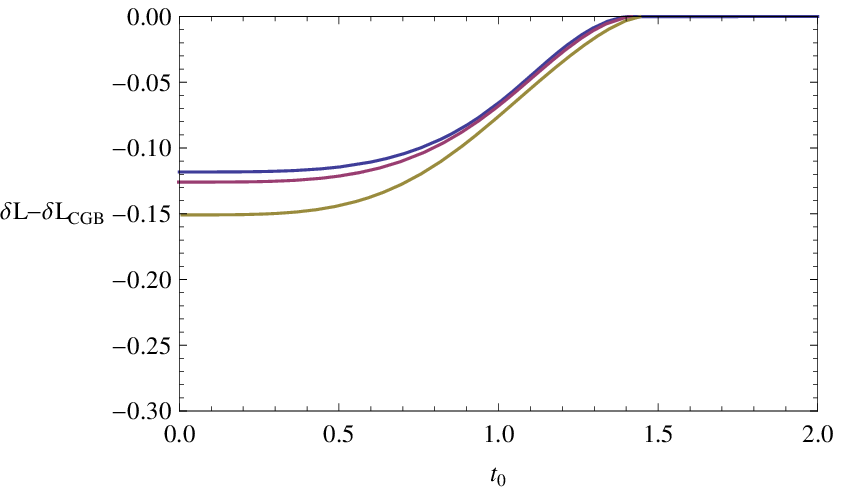}
\label{shell1_RN} }
\subfigure[$\alpha=0.0001$]{
\includegraphics[scale=0.55]{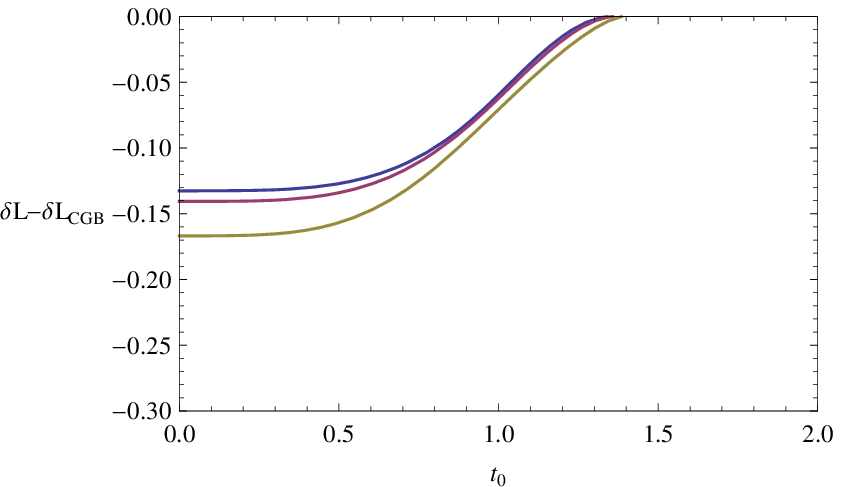}
\label{shell2_RN}
}
\subfigure[$\alpha=0.08$]{
\includegraphics[scale=0.55]{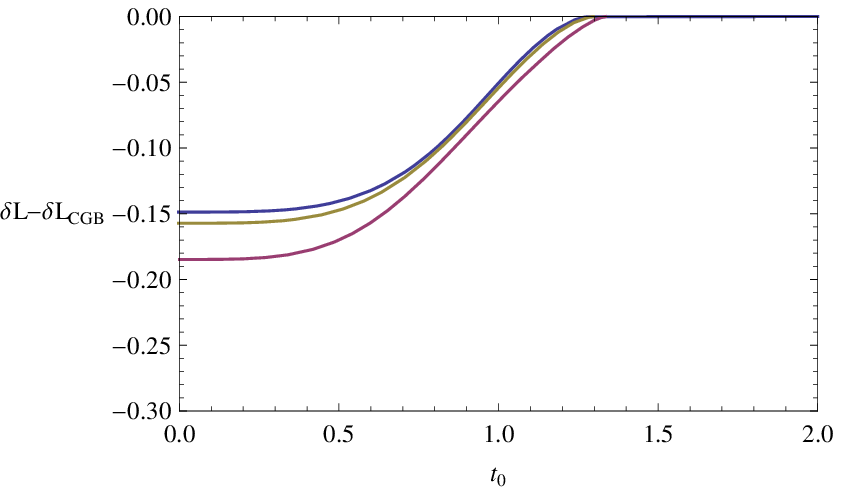}
\label{shell3_RN} }
 \caption{\small Thermalization of the renormalized geodesic lengths in a charged  Gauss-Bonnet Vaidya AdS  black brane for different charge  $Q$ at a fixed Gauss-Bonnet coefficient $\alpha$. The separation of the
boundary field theory operator pair is $\tilde{\ell}=3$.
 The green line, red line and purple line correspond to  $Q=0.00001, 0.5, 1$  respectively.} \label{figa21}
\end{figure}

\begin{figure}
\centering
\subfigure[$Q=0.00001$]{
\includegraphics[scale=0.55]{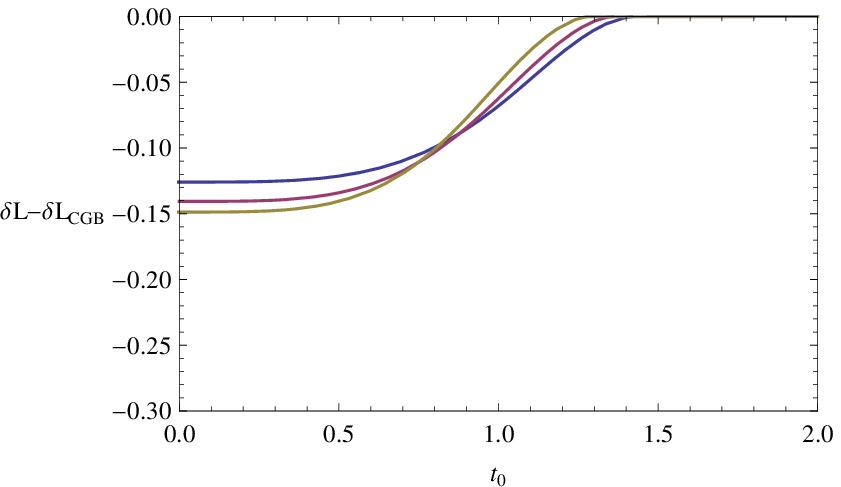}
\label{shell1_RN} }
\subfigure[$Q=0.5$]{
\includegraphics[scale=0.55]{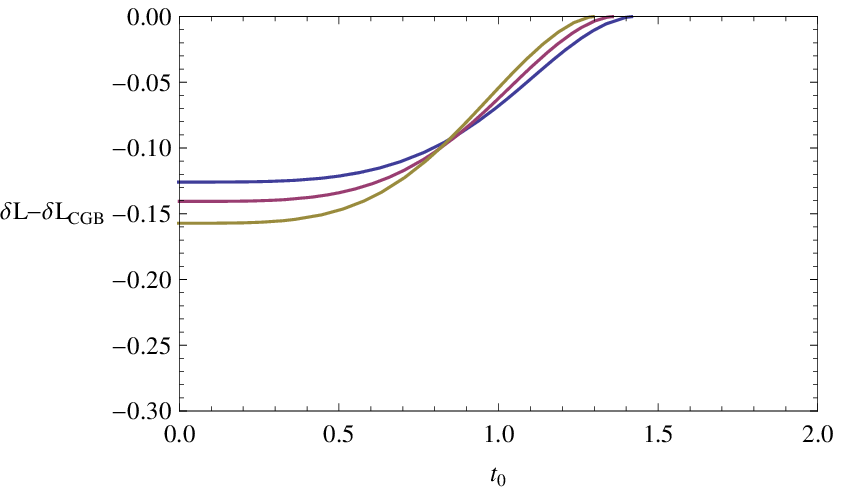}
\label{shell2_RN}
}
\subfigure[$Q=1$]{
\includegraphics[scale=0.55]{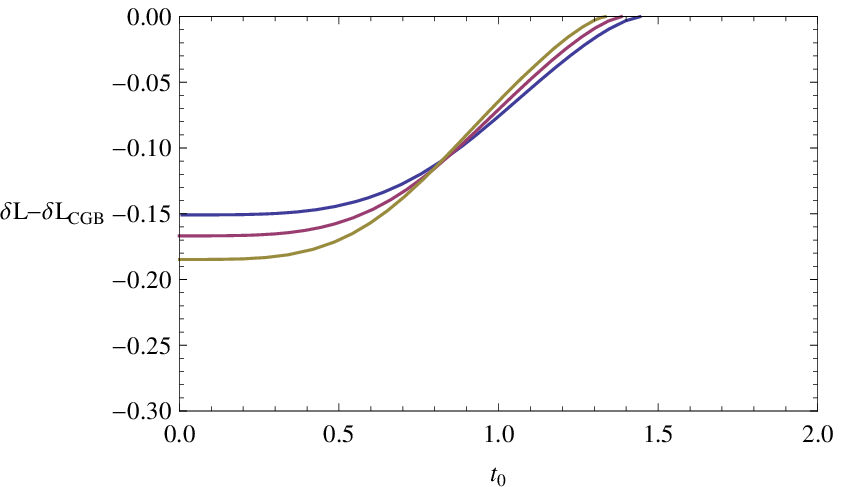}
\label{shell3_RN} }
 \caption{\small Thermalization of the renormalized geodesic lengths in a charged  Gauss-Bonnet Vaidya AdS
  black brane for different Gauss-Bonnet coefficients $\alpha$ at a fixed  charge $Q$. The separation of the
boundary field theory operator pair is $\tilde{\ell}=3$.
 The green line, red line and purple line correspond to  $\alpha=-0.1, 0.0001, 0.08$ respectively.} \label{figa3}
\end{figure}

 Having the numerical result of $z(x)$, we can study
the  renormalized geodesic length  with the help of Eq.(\ref{lren}) and Eq.(\ref{r2}). As done in \cite{GS}, we compare $\delta \tilde{L}$
 at each time with the final values $\delta \tilde{L}_{CGB}$, obtained in a static charged Gauss-Bonnet  AdS black
  brane, \emph{i.e.} $M(\mu)=M$, $Q(\mu)=Q$.
  In this case, the thermalized state  is labeled by the zero point of the vertical coordinate in each picture.
 To get an observable quantity that is $\tilde{l}$ independent, we will plot the quantity $\delta L=\delta \tilde{L}/\tilde{l}$.
Figure (\ref{figa21}) gives the relation between the  renormalized geodesic length and thermalization time for different charge at a fixed Gauss-Bonnet coefficient. In each picture, the vertical axis
indicates the renormalized geodesic length while the horizontal axis indicates the thermalization  time $t_0$.
  For a fixed  Gauss-Bonnet coefficient, \emph{e.g.} $\alpha= 0.08$,  the thermalization time increases as
$Q$ raises. This phenomenon has been also observed previously when we study the motion profile of the geodesic.  In \cite{GS},
the effect of charge on the thermalization time is investigated in Einstein gravity, it was shown that there is an enhancement
of the thermalization time as the ratio of chemical
potential over temperature  increases. Obviously, in  the  Gauss-Bonnet gravity, this phenomenon is not changed. In addition,
 From the  same color line, \emph{e.g.}
green line, in (a), (b) and (c) in Figure (\ref{figa21}),  we know that as the Gauss-Bonnet coefficient increases, the
thermalization time decreases for a fixed charge.
In  Figure (\ref{figa3}), we  plot this graphics for different charge.   Note that \cite{Zeng2013} has investigated this phenomenon for the case $Q=0$.
 It was found  that for a fixed boundary separation there is always a time range in which  the  renormalized geodesic length takes the same value nearly. That is, during that time range,
the Gauss-Bonnet coefficient has little effect on the   renormalized geodesic length. Obviously, (a) in Figure (\ref{figa3}) is consistent with their result. For $Q=0.5$ and  $Q=1$, we also can observe this phenomenon, which are plotted in (b) and (c) in Figure (\ref{figa3}).

\begin{figure}
\centering
\subfigure[$\alpha=-0.1$]{
\includegraphics[scale=0.55]{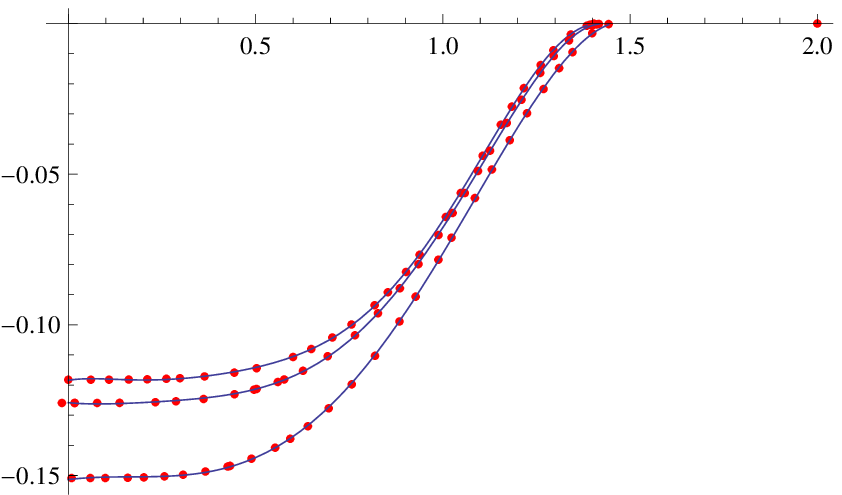}
\label{shell1_RN} }
\subfigure[$\alpha=0.0001$]{
\includegraphics[scale=0.55]{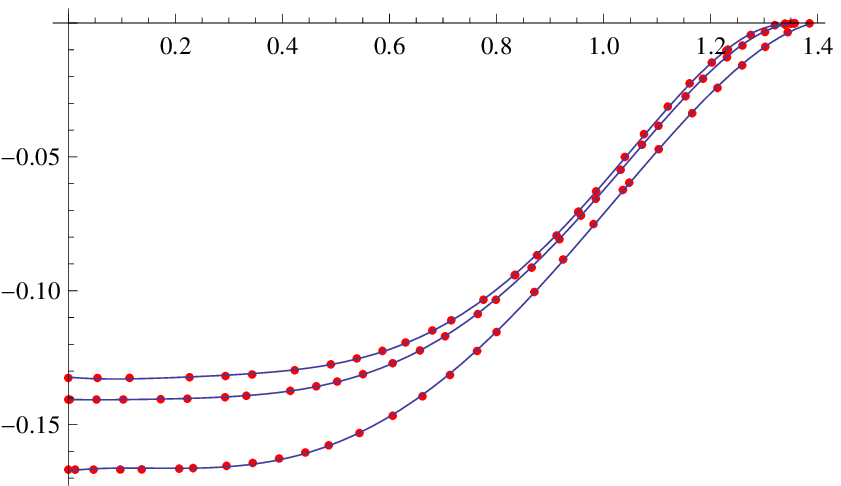}
\label{shell2_RN}
}
\subfigure[$\alpha=0.08$]{
\includegraphics[scale=0.55]{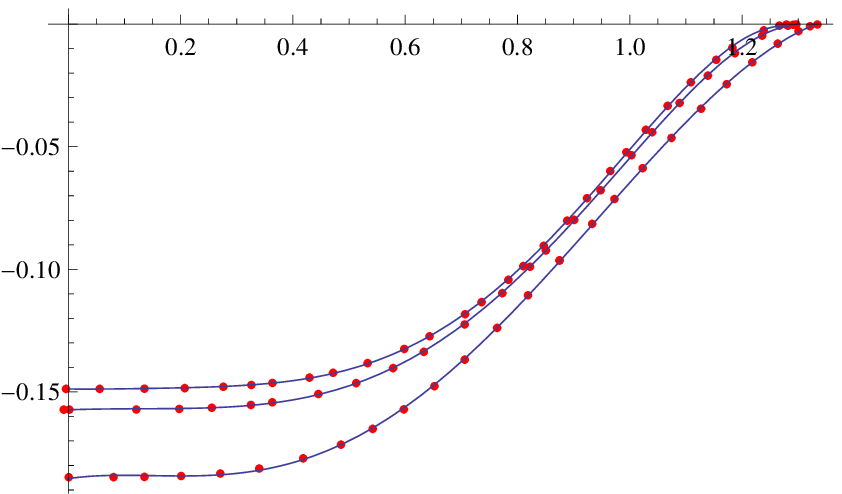}
\label{shell3_RN} }
 \caption{\small Comparison of the function in Eq.(4.3) with the numerical result in Figure 3.} \label{figa4}
\end{figure}

\begin{figure}
\centering
\subfigure[$\alpha=0.0001$ and $\tilde{\ell}=3$]{
\includegraphics[scale=0.75]{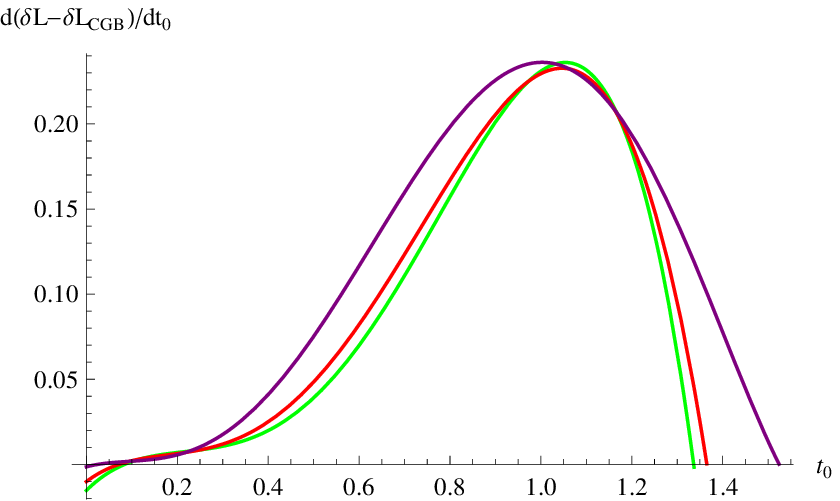}
\label{}
}
\subfigure[$Q=1$ and $\tilde{\ell}=3$]{
\includegraphics[scale=0.75]{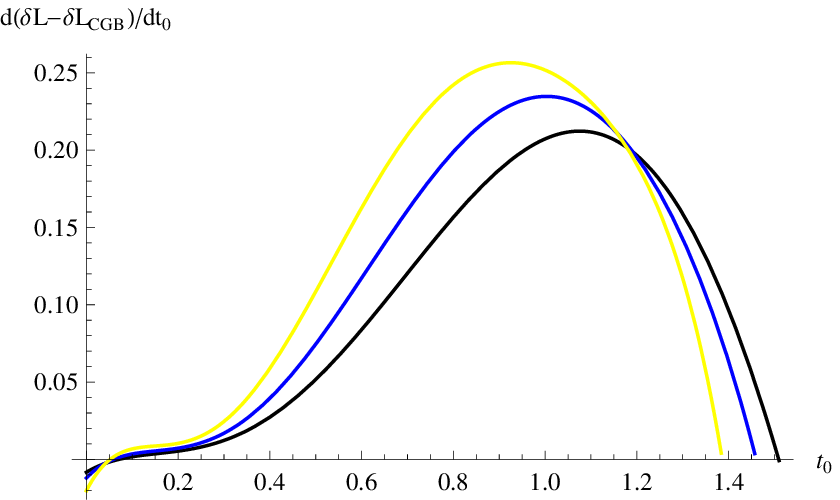}
\label{} }
 \caption{\small Thermalization velocity of the renormalized geodesic lengths in a charged  Gauss-Bonnet Vaidya AdS  black brane.
 The green line, red line and purple line in (a) correspond to  $Q=0.00001, 0.5, 1$ and the black line, blue line and yellow line
 in (b) correspond to   $\alpha=-0.1, 0.0001, 0.08$
 respectively.} \label{fignew1}
\end{figure}

Interestingly, we find the thermalization curve for a fixed charge and  Gauss-Bonnet coefficient in Figure (\ref{figa21})
 can be fitted as a function of $t_0$. Figure (\ref{figa4}) is the comparison result of the numerical  curve and function curve.
  At a fixed $\alpha$, one can get the function of the thermalization curve for different charge.
  For example, at $\alpha=0.0001$, the thermalization curve for $Q=0.00001, 0.5, 1$ can be
   expressed respectively as \footnote{For higher order power of $t_0$, we find it has few contributions
    to the thermalization, including the phase transition point which will be discussed next. }
\begin{equation}
  \begin{cases}
   g_1=-0.132319-0.015338 t_0+0.132916 t_0^2-0.393653 t_0^3+0.560646 t_0^4-0.190989 t_0^5-0.0210284 t_0^6 \\
   g_2=-0.140525-0.0101622 t_0+0.101997 t_0^2-0.348701 t_0^3+0.626571 t_0^4-0.324671 t_0^5+0.033151 t_0^6 \\
 g_3=-0.166815-0.00146773 t_0+0.0355258 t_0^2-0.209675 t_0^3+0.691328 t_0^4-0.54898 t_0^5+0.129204 t_0^6
  \end{cases}
  \end{equation}
For small time, the function is determined  by the lower power of $t_0$,
while for large time it  is determined by the higher  power  of $t_0$. With the function, we can get the thermalization
velocity, which is plotted in   Figure (\ref{fignew1}). From this figure, we can observe two interesting phenomena for the
thermalization process. One is that the thermalization velocity is negative at the initial time while it is positive at
 the middle and
later periods, which means that the thermalization
process  is non-monotonic. In fact, this non-monotonic behavior has also  been observed in  \cite{CK}.
The reason  for this behavior is that at the initial time the thermalization is ``quantum"  while at the
later periods, it is
``classical".  The author in  \cite{CK} further argued  that at the ``quantum"  stage   the slope of the thermalization curve, namely
the thermalization velocity, is negative and at the ``classical" stage the slope is positive. Obviously, our result plotted in
Figure (\ref{fignew1}) confirms their argument.
The other is that there is a phase transition point for the thermalization velocity, which divides the thermalization into an
acceleration phase and a deceleration phase. (a) and (b) in Figure (\ref{fignew1}) represent respectively the effect of $Q$ and
$\alpha$ on the phase transition points. From both figures, we know that the phase transition points vary for different  $Q$ and
$\alpha$. Especially in (b), we can observe clearly that as $\alpha$ increases, the phase transition points shift left.
At a fixed  $Q$ and
$\alpha$, we can get the value of the phase transition point.
For the case $Q=0.5$ and  $\alpha=0.0001$,
it is easy to find that in the time range
$0<t_0<1.04978$, the thermalization is an acceleration process while for $t_0>1.04978$ it is a deceleration
 process before it approaches to the equilibrium state. Surely  as the slope of this velocity curve is produced, we also can get
 the values of the acceleration and
deceleration.

%======================figure1===================

\begin{table}
\begin{center}\begin{tabular}{|l|c|c|c|c|c|c|}
 %\MC{3}{c}{\text{caption}}\\[5pt]
 \hline
% & \multicolumn{3}{c||}{MGCDM}   & \multicolumn{3}{c}{$\Lambda$CDM}  \\ \hline
%                             &        MGCDM        &                  &             &      $\Lambda$CDM    &                   & \\ \hline
% \MC{3}{|c|c|}{\ZZ{-8pt}{15pt}\hfill\normalsize   \hfill  \hfill\normalsize MGCDM     \hfill\normalsize $\Lambda$CDM  }\\ \hline
% \ZZ{-6pt}{22pt}

          & \multicolumn{3}{|c|}   { $v_{\star}$=-0.856}     &      \multicolumn{3}{|c|}  { $v_{\star}$= -0.456} \\ \hline
&  $\alpha$=-0.1       &$\alpha$=0.0001  &$\alpha$=0.08   &  $\alpha$=-0.1       &$\alpha$=0.0001  &$\alpha$=0.08  \\  \hline
$Q$=0.00001     &0.691064      &  0.625561    & 0.560275  & 1.01538      &  0.949617   &  0.887499   \\ \hline
$Q$=0.5         &0.690223        & 0.624786    & 0.559595   &1.02213       & 0.958193     & 0.897744    \\ \hline
$Q$=1           &0.687522       & 0.622449      & 0.557604   &1.03968       & 0.981075      & 0.925744    \\ \hline
\end{tabular}
\end{center}
\caption{The thermalization time $t_0$ of the geodesic probe for different Gauss-Bonnet
 coefficient $\alpha$ and different charge $Q$ at $v_{\star}=-0.856,-0.456$ respectively. }\label{tab:g1}
\end{table}
%======================figure2====================

\begin{figure}
\centering
\subfigure[$Q=0.00001, \alpha=-0.1$]{
\includegraphics[scale=0.51]{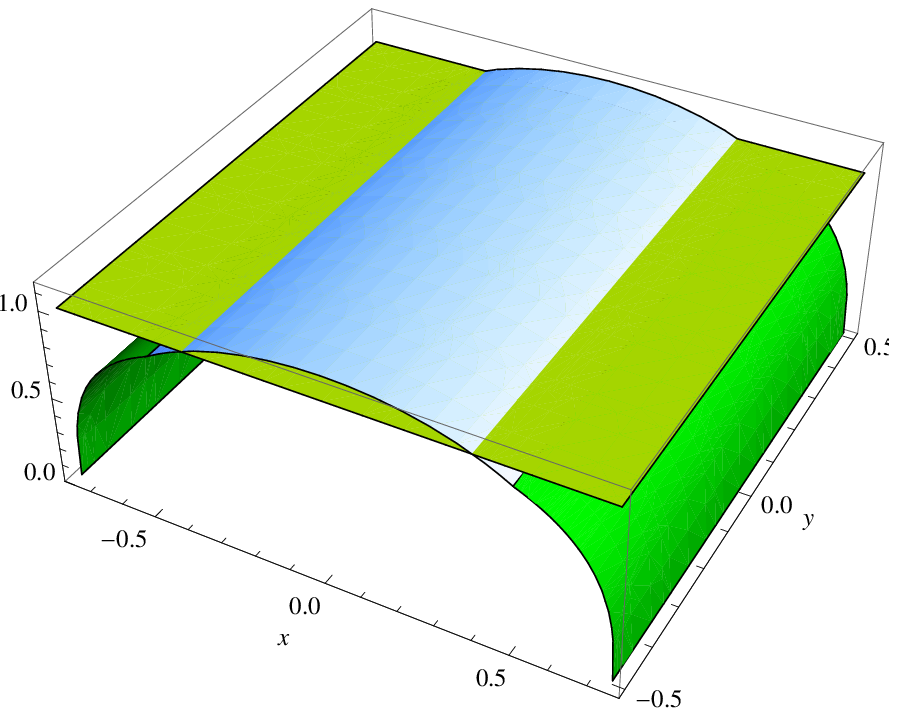}
}
\subfigure[$Q=0.00001, \alpha=0.0001$]{
\includegraphics[scale=0.51]{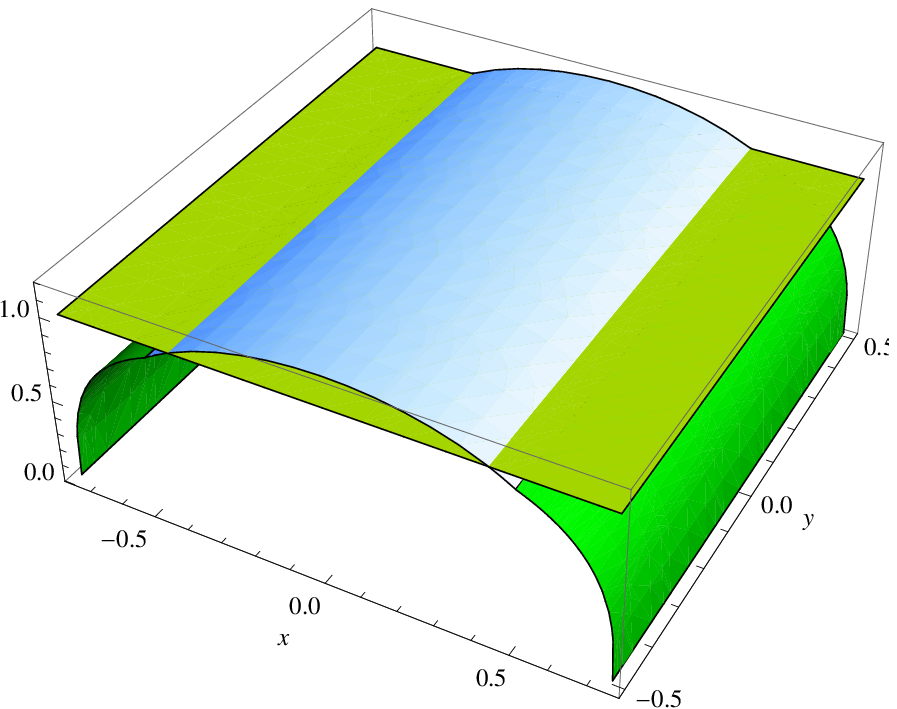}
}
\subfigure[$Q=0.00001, \alpha=0.08$]{
\includegraphics[scale=0.51]{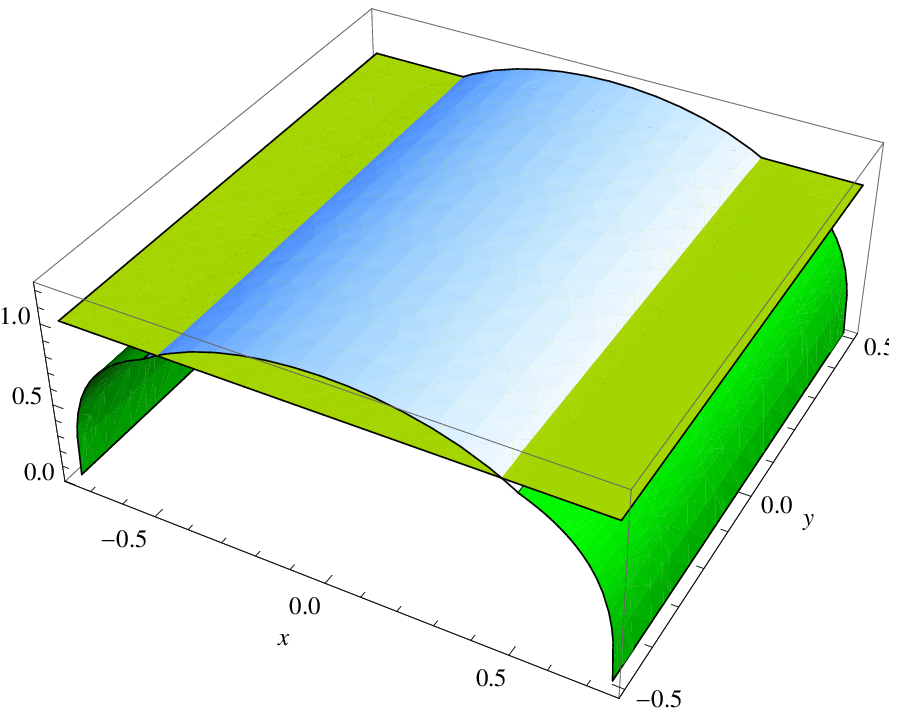}
}
\subfigure[$Q=0.5, \alpha=-0.1$]{
\includegraphics[scale=0.51]{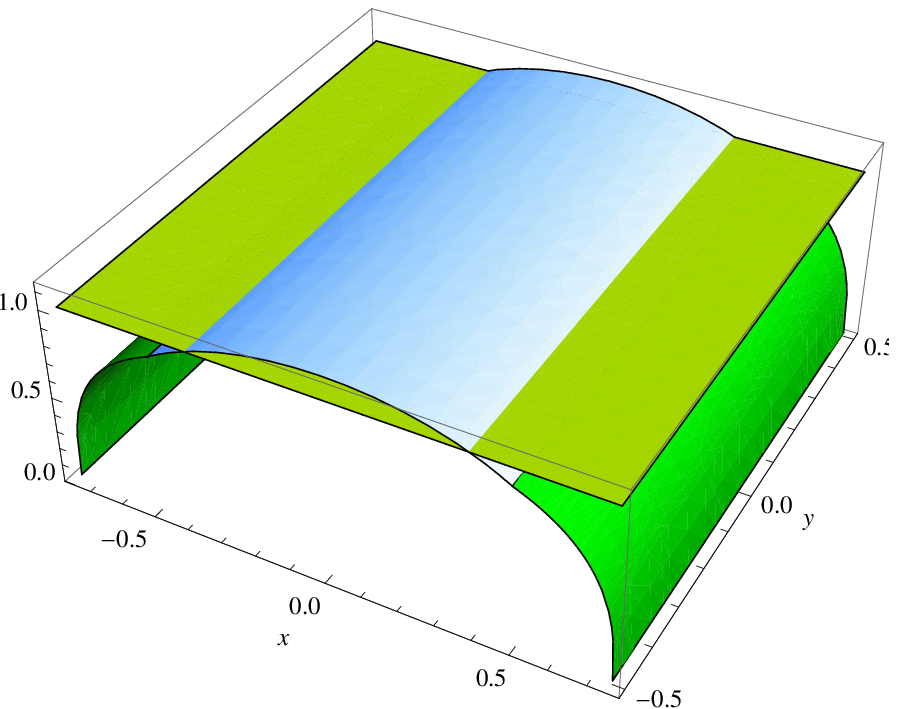}
 }
\subfigure[$Q=0.5, \alpha=0.0001$]{
\includegraphics[scale=0.51]{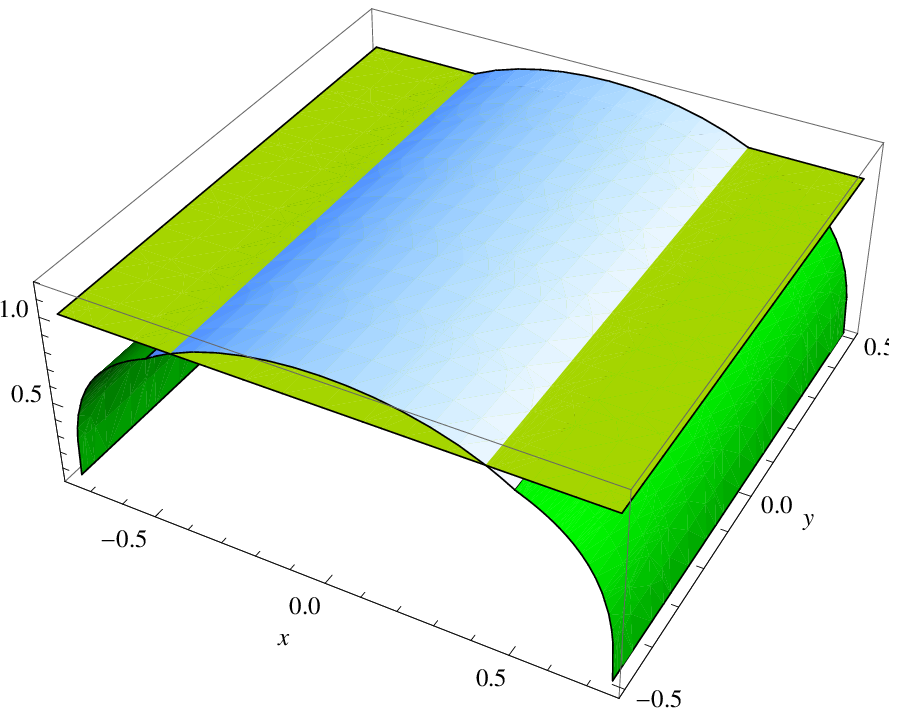}
}
\subfigure[$Q=0.5, \alpha=0.08$]{
\includegraphics[scale=0.51]{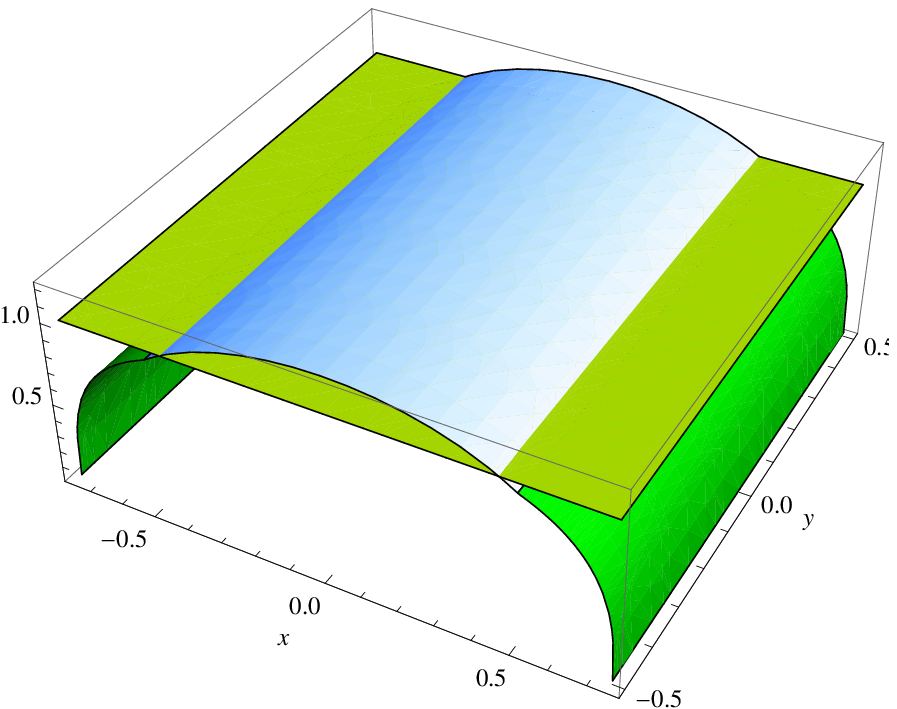}}
\subfigure[$Q=1, \alpha=-0.1$]{
\includegraphics[scale=0.51]{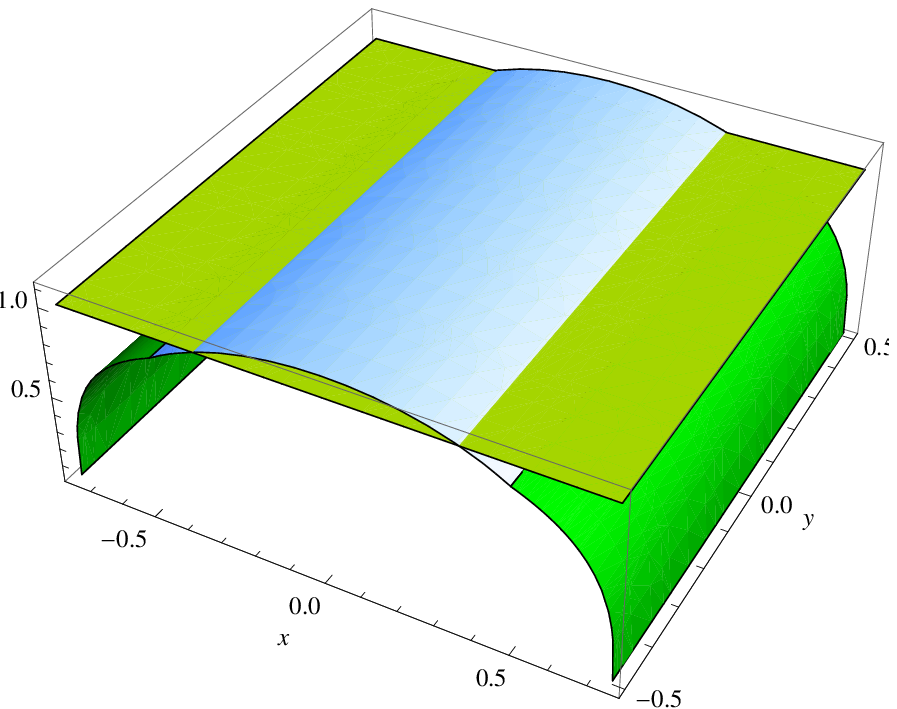} }
\subfigure[$Q=1, \alpha=0.0001$]{
\includegraphics[scale=0.51]{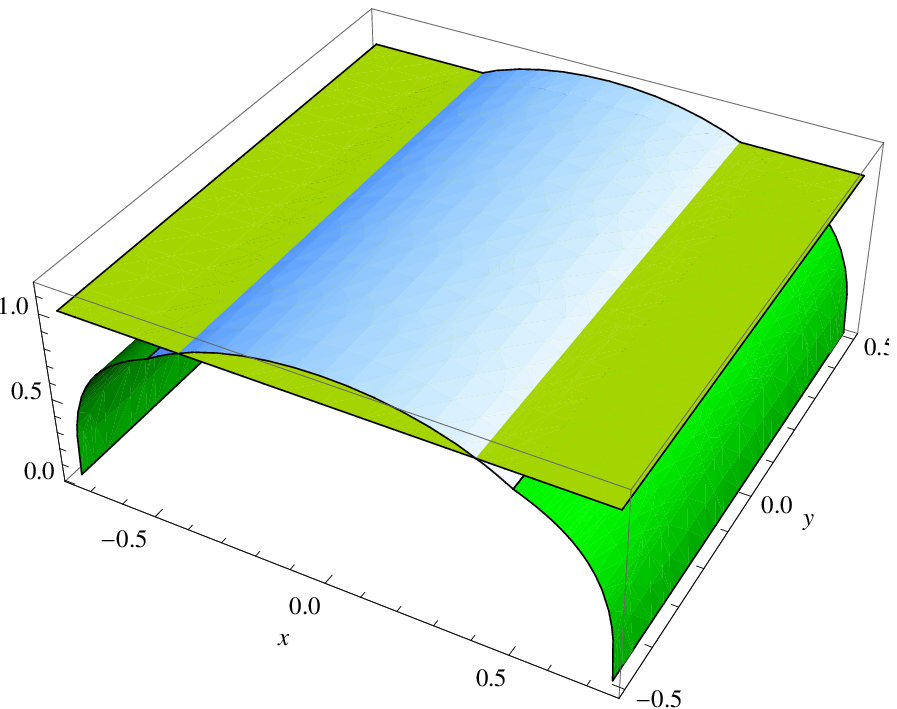}
}
\subfigure[$Q=1, \alpha=0.08$]{
\includegraphics[scale=0.51]{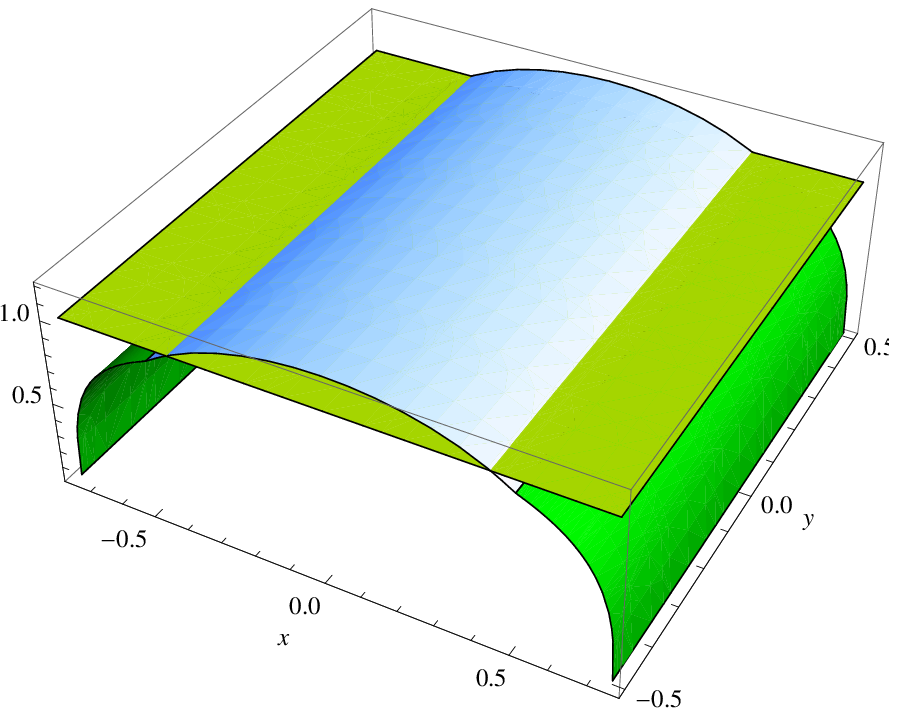}

 }
 \caption{\small Motion profile of the minimal area   in the charged Gauss-Bonnet Vaidya AdS black brane.
  The boundary separation along the $x$ direction is $1.5$, and along the $y$ direction is $1$, the initial
  time is $v_*=-0.252$. The yellow surface is the location of the horizon.  The  position of the shell  is described by the junction between the  white surface and the green surface.} \label{figa11}
\end{figure}

Adopting similar strategy, we also can study the motion profile of minimal area  as well as  the  change of
 the renormalized minimal  area surface  to thermalization
 time. Based on the motion equations in (\ref{aequation}) and the boundary conditions in  (\ref{initial}),
the numerical solution of $z(x)$  can be produced. In this case, we can get the  motion profile of minimal area  for different  charge $Q$ and Gauss-Bonnet coefficient $\alpha$,
which is shown in Figure (\ref{figa11}). From  this figure, we know that for a fixed charge, e.g.  the first row,
as the Gauss-Bonnet coefficient increases, the shell surface approaches
 to the horizon surface step by step, which means the thermalization is faster. For a fixed Gauss-Bonnet coefficient, e.g.
 the third column,
as the charge increases, the shell surface  is removed from the horizon surface step by step, which means the thermalization
is slower.
 The thermalization time for different $\alpha$ and $Q$ have been listed in Table (\ref{tabaa1}). It is shown  that
  for a fixed charge the thermalization time decreases
 as $\alpha$ becomes larger, while for a fixed $\alpha$,  the thermalization time increases
 as $Q$ becomes larger. That is, the charge has an inverse effect compared with the  Gauss-Bonnet coefficient
  on the thermalization time.
  This phenomenon is similar to that of the geodesics.

%======================figure1====================

\begin{table}
\begin{center}\begin{tabular}{|l|c|c|c|}
 %\MC{3}{c}{\text{caption}}\\[5pt]
 \hline
% & \multicolumn{3}{c||}{MGCDM}   & \multicolumn{3}{c}{$\Lambda$CDM}  \\ \hline
%                             &        MGCDM        &                  &             &      $\Lambda$CDM    &                   & \\ \hline
% \MC{3}{|c|c|}{\ZZ{-8pt}{15pt}\hfill\normalsize   \hfill  \hfill\normalsize MGCDM     \hfill\normalsize $\Lambda$CDM  }\\ \hline
% \ZZ{-6pt}{22pt}
%   &      \multicolumn{3}{|c|}  { $v_{\star}$= -0.456} \\ \hline
&$\alpha$=-0.1       &$\alpha$=0.0001  &$\alpha$=0.08  \\  \hline
$Q$=0.00001        &1.01732      &  0.963053   & 0.911401     \\ \hline
$Q$=0.5           &1.02170       & 0.968943  & 0.918953     \\ \hline
$Q$=1            &1.03334       &0.984877      & 0.939824      \\ \hline
\end{tabular}
\end{center}
\caption{The thermalization time $t_0$ of the geodesic probe for different Gauss-Bonnet coefficient $\alpha$ and
 different charge $Q$ at a initial time $v_{\star}=-0.252$}\label{tabaa1}.
\end{table}

%======================figure1====================

%======================figure1====================

\begin{figure}
\centering
\subfigure[$\alpha=-0.1$]{
\includegraphics[scale=0.55]{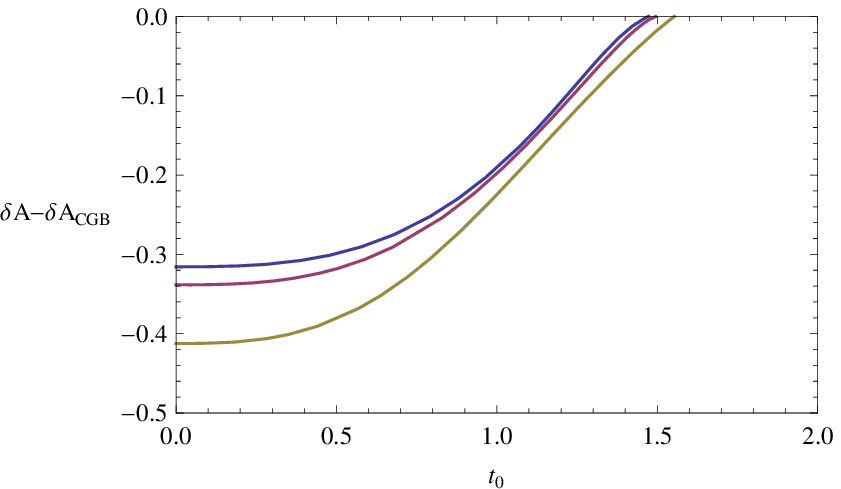}
\label{shell1_RN} }
\subfigure[$\alpha=0.0001$]{
\includegraphics[scale=0.55]{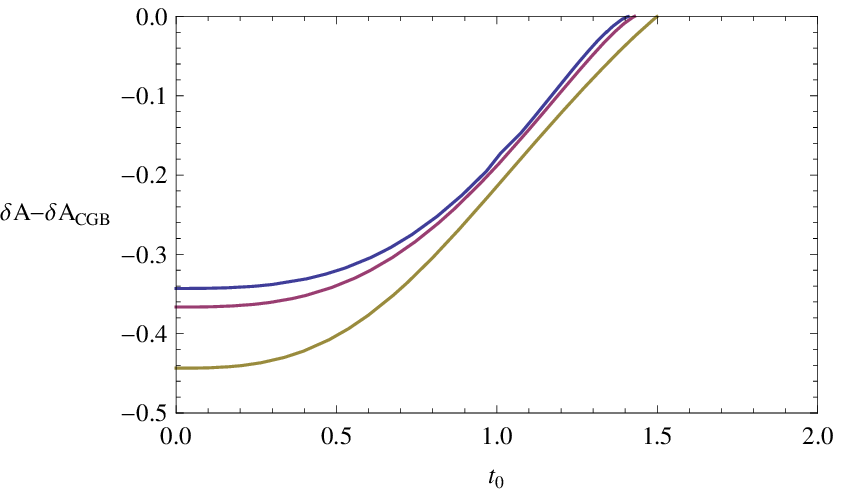}
\label{shell2_RN}
}
\subfigure[$\alpha=0.08$]{
\includegraphics[scale=0.55]{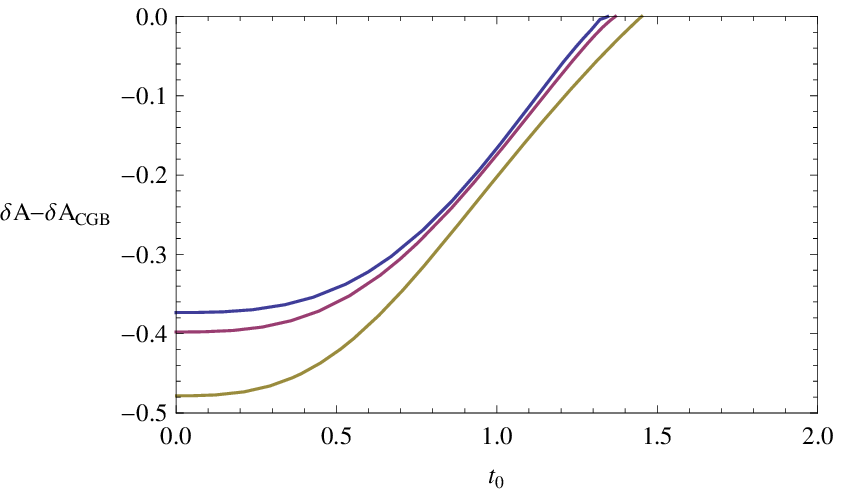}
\label{shell3_RN} }
 \caption{\small Thermalization of the renormalized minimal surface  area  in a charged  Gauss-Bonnet Vaidya AdS black brane  for different charge $Q$ at a fixed Gauss-Bonnet coefficients $\alpha$. The separation of the
boundary field theory operator pair is $\tilde{\ell}=2$.
 The green line, red line and purple line correspond  to  $Q=0.00001, 0.5, 1$ respectively.} \label{figa33}
\end{figure}

\begin{figure}
\centering
\subfigure[$Q=0.00001$]{
\includegraphics[scale=0.55]{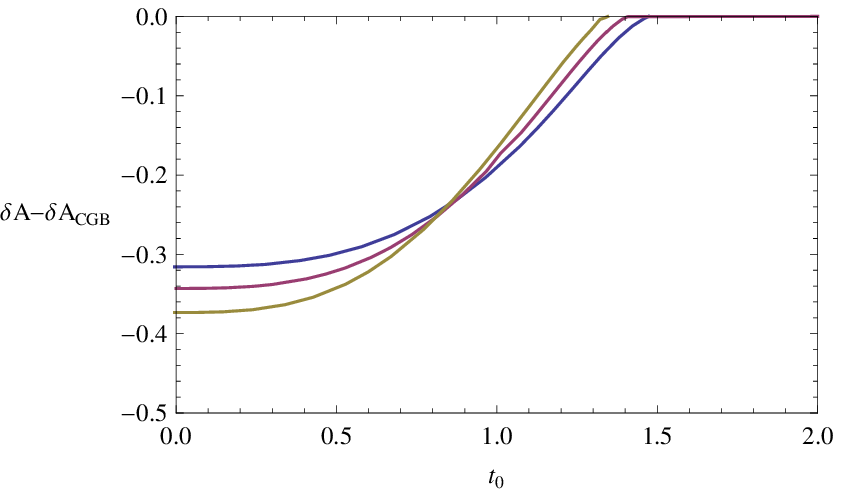}
\label{shell1_RN} }
\subfigure[$Q=0.5$]{
\includegraphics[scale=0.55]{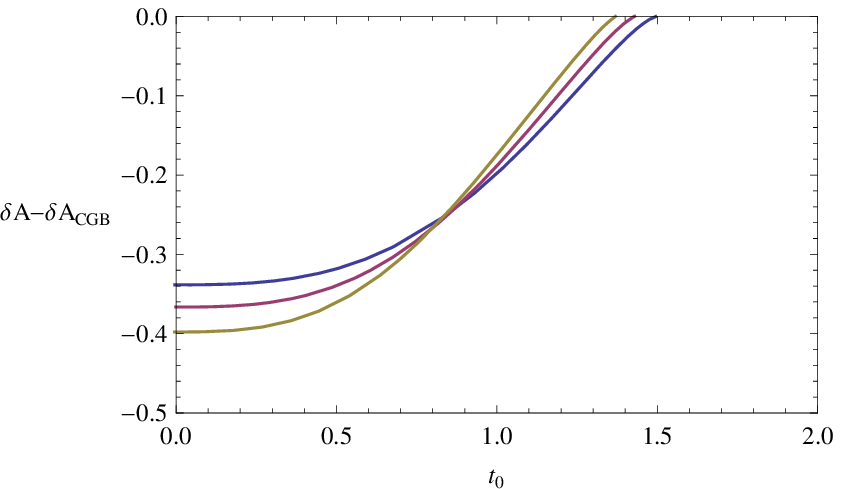}
\label{shell2_RN}
}
\subfigure[$Q=1$]{
\includegraphics[scale=0.55]{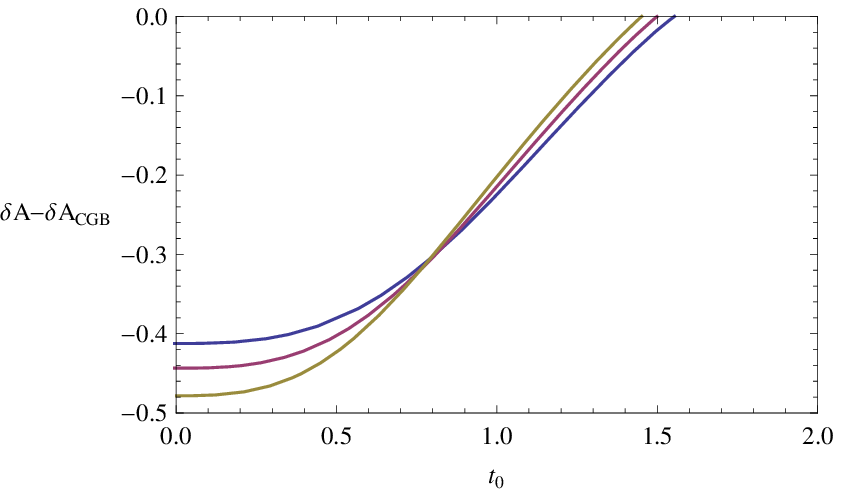}
\label{shell3_RN} }
 \caption{\small Thermalization of the renormalized minimal surface  area  in a charged Gauss-Bonnet Vaidya AdS black brane  for different Gauss-Bonnet coefficients $\alpha$ at a fixed charge $Q$. The separation of the
boundary field theory operator pair is $\tilde{\ell}=2$.
 The green line, red line and purple line correspond  to  $\alpha=-0.1, 0.0001, 0.08$ respectively.} \label{figa34}
\end{figure}

\begin{figure}
\centering
\subfigure[$\alpha=-0.1$]{
\includegraphics[scale=0.55]{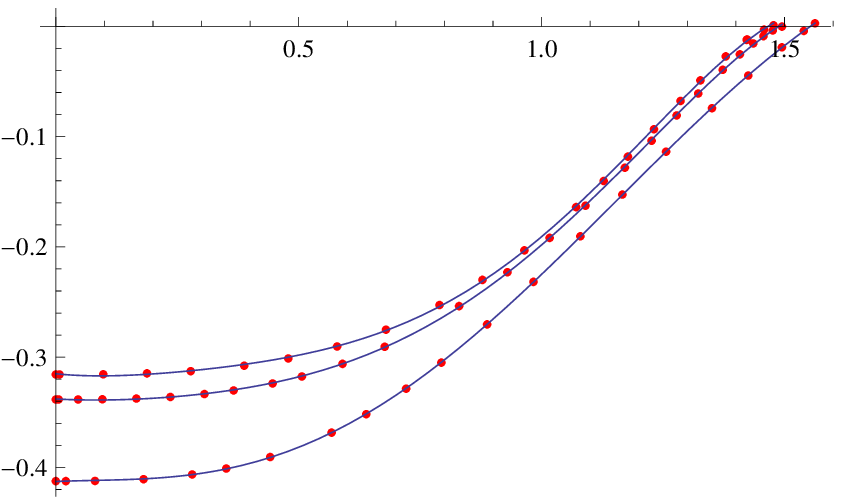}
\label{shell1_RN} }
\subfigure[$\alpha=0.0001$]{
\includegraphics[scale=0.55]{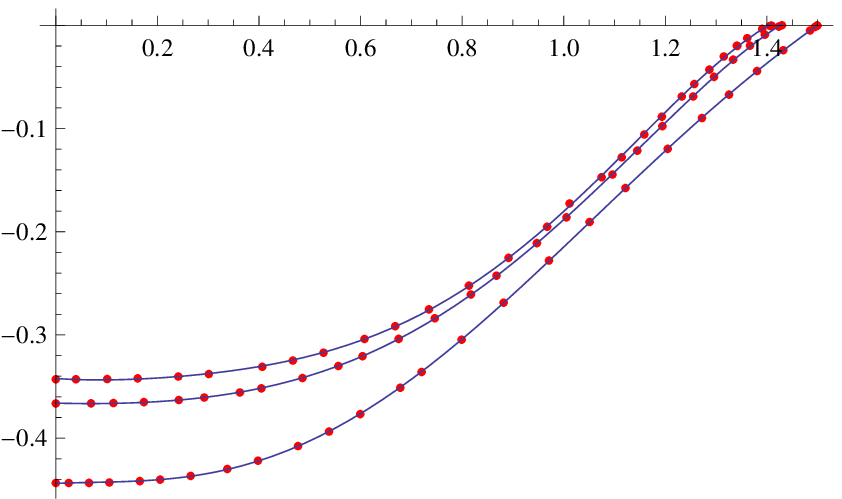}
\label{shell2_RN}
}
\subfigure[$\alpha=0.08$]{
\includegraphics[scale=0.55]{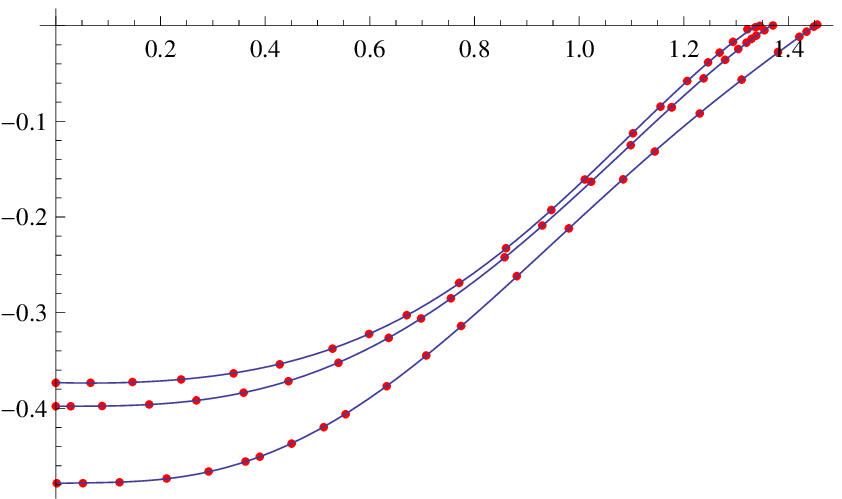}
\label{shell3_RN} }
 \caption{\small Comparison of the function in Eq.(4.4) with the numerical result in Figure 8. } \label{figa35}
\end{figure}

\begin{figure}
\centering
\subfigure{
\includegraphics[scale=0.95]{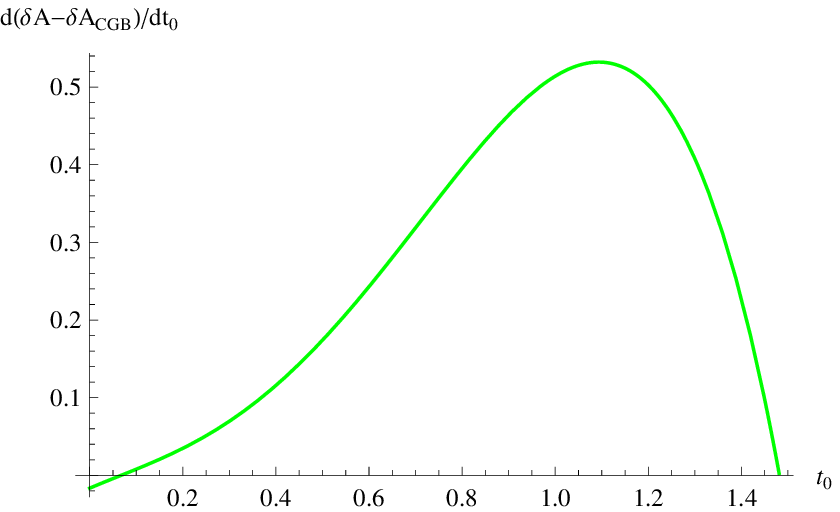} }
 \caption{\small Thermalization velocity of the renormalized minimal surface  area  in a charged  Gauss-Bonnet Vaidya AdS  black brane
for $Q=0.00001$ and $\alpha= 0.08$.} \label{fignew3}
\end{figure}

Substituting the numerical result
  of $z(x)$ into  (\ref{aren}), we can get the renormalized minimal  area surface. Similar to the case of geodesic, we will plot $\delta A-\delta A_{CGB}$,
   where  $\delta A=\delta \tilde{A}/\tilde{l}$ and $\delta A_{CGB}$ is the  renormalized minimal  area surface    for a  charged Gauss-Bonnet AdS black brane.
   The relation between the renormalized minimal  area surface    and thermalization time for different charge $Q$
 is given in Figure  (\ref{figa33}) for a fixed  Gauss-Bonnet coefficient $\alpha$, in which the vertical axis indicates
 the renormalized minimal  area surface while the horizontal axis indicates
 the thermalization time $t_0$.  For a fixed  $\alpha$, we find larger the charge $Q$ is, longer the thermalization time is.
 That is to say, as the chemical potential in the dual field theory increases, the thermalization time rises too,
 which is the same as that obtained by studying the motion profile of the minimal  area.  In addition, for a fixed charge, we also
 study the effect of the Gauss-Bonnet coefficient on the thermalization time, which is shown in  Figure (\ref{figa34}).
It is obvious that larger the Gauss-Bonnet coefficient is,  shorter the thermalization time is, which means that
  the quark gluon plasma is easier to thermalize. This behavior
is similar to that of the geodesic which is given in Figure (\ref{figa21}).
 As the case of the renormalized geodesic length, we find
 there is also an overlapped region for the case $Q=0.00001, 0.5,1$ respectively in Figure  (\ref{figa34}).
 We also can get the functions of the renormalized minimal  area surface with respect to the thermalization time.
  At $\alpha=0.08$, the functions of the thermalization curve for $Q=0.00001, 0.5, 1$ can be expressed respectively as
\begin{equation}
  \begin{cases}
   h_1=-0.373032-0.0165426 t_0+0.128311 t_0^2-0.0886585 t_0^3+0.379981 t_0^4-0.19601 t_0^5  \\
   h_2=-0.397816+0.000718227 t_0-0.0130504 t_0^2+0.383351 t_0^3-0.115431 t_0^4-0.0324124 t_0^5 \\
 h_3=-0.478534+0.0172206 t_0-0.165658 t_0^2+1.09677 t_0^3-0.887269 t_0^4+0.215318 t_0^5
  \end{cases}
  \end{equation}
 That is, the thermalization curve can be described by a function
of time $t_0$ with different modulus for different charges and different  Gauss-Bonnet coefficients.
With the thermalization curve, we further can get the thermalization velocity.  For the
 case $\alpha=0.08$ and $Q=0.00001$, the thermalization velocity is plotted in  Figure (\ref{fignew3}). From this figure, we
 also can observe the non-monotonic behavior and phase transition point of the thermalization process, which is similar to that of
 the  renormalized geodesic length.
We also can obtain the acceleration phase and deceleration phase for this case. By the slope of the thermalization velocity,
we know that the acceleration
range is $0<t_0<1.106$ and  the deceleration range is $1.106<t_0<1.481$. Adopting the same strategy, we also
 can get the phase transition points for other $Q$  and $\alpha$.

%=================================================
%======================figure2====================
%======================figure2====================

\section{Conclusions}

Effect of the chemical potential and  correction parameter on the thermalization in the dual boundary field theory is investigated by
considering  the collapse of a shell of charged dust that  interpolates between a pure AdS and a
 charged  Gauss-Bonnet AdS black brane.
   The two-point function and expectation values of Wilson loop are chosen as the thermalization probes, which
    are dual to the  renormalized geodesic length and minimal  area surface   in the bulk.
We  first study the motion profiles of the geodesic and minimal surface   and find that larger the
 Gauss-Bonnet coefficient is,  shorter the  thermalization time is, and  larger the charge is,
  longer the thermalization time is. At the initial stage of the thermalization, we find that the
  charge has little effect on the thermalization time. We reproduce this result by studying the relation between the  renormalized
geodesic length and time as well as the renormalized minimal surface  area and time respectively.

 In addition, we also find the functions of the thermalization probes with respect to the thermalization time by
 fitting the numerical result. Though this is a
 naive test, we still  can get some useful information. Firstly, we get the thermalization velocity for
 a fixed  charge and  Gauss-Bonnet coefficient. From the velocity curve, on one hand, we know that the  thermalization process is
 non-monotonic, and on the other hand  we find there is a phase transition point, which divides the
   thermalization into
   an acceleration phase and a deceleration phase. Secondly according to the slope of the velocity
 curve, we also get the acceleration and deceleration of the thermalization.
Thus our investigation
   provides a more accurate description of the thermalization process.
 Note that in our naive test, we can not see how the charge and Gauss-Bonnet coefficient affect
   the thermalization quantificationally since the coefficients of the functions are not  determined. In future, we expect to
   find an analytical formulism to study the thermalization so that we can get more useful information on the thermalization process.

\section*{Acknowledgements}

We would like to give great thanks to the anonymous referee for his helpful suggestions about this paper.  Xiao-Xiong Zeng would like to thank Hongbao Zhang for his encouragement and various valuable suggestions during this work.
 This work is supported in part by the National
 Natural Science Foundation of China (Grant Nos. 11365008, 61364030). It is also supported by the Natural Science
Fund of Education Department of Hubei Province (Grant No. Q20131901)

\section*{Appendix: Derivation of the Hamiltonian from the Lagrangian}

Suppose that there exists a Hamiltonian $H$ such that
\begin{equation}
U(x,y;T)=\langle x|e^{-iHT}|y\rangle=\mathcal{N}\prod_\tau \int
d^{d+1}X(\tau)\sqrt{-g(\tau)}e^{i\int_0^Td\tau L(\tau)}
\end{equation}
with $X(0)=y$, $X(T)=x$, and the Lagrangian
$L=\frac{1}{2}(g_{\mu\nu}
\frac{dX^\mu}{d\tau}\frac{dX^\nu}{d\tau}-m^2)$.
Then we have
\begin{equation}
i\frac{\partial}{\partial T}U(x,y;T)=HU(x,y;T),
\end{equation}
and
\begin{eqnarray}\label{Peskin}
U(x,y;T)&=&N\int d^{d+1}X\sqrt{-g(X)}\nonumber\\
&&e^{\frac{i}{2\epsilon}g_{\mu\nu}(\frac{x+X}{2})(x-X)^\mu(x-X)^\nu-\frac{i}{2}\epsilon
m^2} U(X,y;T-\epsilon),
\end{eqnarray}
where $\epsilon$ is a small quantity, to be taken to go to zero in
the later calculation. Now for convenience, we would like to resort
to the Rienmann normal coordinate at $x$, where the behavior of
metric is simplified as
\begin{equation}
g_{\mu\nu}(x)=\eta_{\mu\nu},\partial_\rho
g_{\mu\nu}(x)=0,\partial_\rho\partial_\sigma
g_{\mu\nu}(x)=-\frac{1}{3}[R_{\mu\rho\nu\sigma}(x)+R_{\mu\sigma\nu\rho}(x)].
\end{equation}
So by Taylor expanding all the involved functions at $x$ and $T$,
Eq.(\ref{Peskin}) gives rise to
\begin{eqnarray}
U(x,y;T)&=&N\int
d^{d+1}X\sqrt{-g(x)}[1-\frac{1}{6}R_{\rho\sigma}(x)(x-X)^\rho(x-X)^\sigma+\cdot\cdot\cdot]\nonumber\\
&&e^{\frac{i}{2\epsilon}g_{\mu\nu}(x)(x-X)^\mu(x-X)^\nu}(1-\frac{i}{2}\epsilon
m^2+\cdot\cdot\cdot)\nonumber\\
&&[1-(x-X)^\rho\partial_\rho+\frac{1}{2}(x-X)^\rho(x-X)^\sigma\partial_\rho\partial_\sigma+\cdot\cdot\cdot]\nonumber\\
&&(1-\epsilon\frac{\partial}{\partial
T}+\cdot\cdot\cdot)U(x,y;T).\nonumber\\
\end{eqnarray}
Here we only keep those terms up to the first order of $\epsilon$
after the Gaussian integral, where the normalization constant $N$ can be fixed by the zero order
equation and $H$ can be determined by the first order equation as
\begin{equation}
H=-\frac{1}{2}[\nabla^a\nabla_a-m^2]+\frac{1}{6}R.
\end{equation}
Similarly, if the Lagrangian is given by
\begin{equation}
L=\frac{1}{2}(g_{\mu\nu}
\frac{dX^\mu}{d\tau}\frac{dX^\nu}{d\tau}-m^2)+\frac{1}{6}R,
\end{equation}
then the corresponding Hamiltonian will be shifted to
\begin{equation}
H=-\frac{1}{2}[\nabla^a\nabla_a-m^2].
\end{equation}

\end{document}